\documentclass[12pt,letterpaper]{article}
\usepackage[affil-it]{authblk}
\usepackage[english]{babel}
\usepackage[utf8]{inputenc}
\usepackage{amsmath}
\usepackage{amssymb}
\usepackage{graphicx}
\usepackage{adjustbox}
\usepackage{xspace}
\usepackage{xcolor}
\usepackage{authblk}
\usepackage{csquotes}
\usepackage{multirow}
\usepackage{physics}
\usepackage{accents}
\newcommand*{\pbar}[1]{\accentset{(-)}{#1}}

\newcommand{\raq}{$r_A^2$}

\usepackage{hyperref}
\usepackage[margin=1.0in]{geometry}
\usepackage[style=numeric-comp,sorting=none]{biblatex}
\bibliography{elementary}

\date{June 1, 2022}

\title{Neutrino Scattering Measurements on Hydrogen and Deuterium: A Snowmass White Paper}

\author[1]{Luis Alvarez-Ruso}
\author[2,3]{Joshua L. Barrow}
\author[4]{Leo Bellantoni}
\author[4]{Minerba Betancourt}
\author[4]{Alan Bross}
\author[5]{Linda Cremonesi}
\author[6]{Kirsty Duffy}
\author[7]{Steven Dytman}
\author[8]{Laura Fields}
\author[9]{Tsutomu Fukuda}
\author[10]{Diego Gonz\'alez-D\'iaz} 
\author[11]{Mikhail Gorchtein}
\author[12,4]{Richard J. Hill} 
\author[4]{Thomas Junk}
\author[13]{Dustin Keller}
\author[14]{Huey-Wen Lin} 
\author[15]{Xianguo Lu} 
\author[14]{Kendall Mahn}
\author[16,17]{Aaron S. Meyer}
\author[4]{Tanaz Mohayai}
\author[4]{Jorge G. Morf\'{i}n}
\author[18]{Joseph Owens}
\author[4]{Jonathan Paley}
\author[19]{Vishvas Pandey}
\author[20]{Gil Paz} 
\author[21]{Roberto Petti}
\author[12,4]{Ryan Plestid} 
\author[4]{Bryan Ramson}
\author[17]{Brooke Russell}
\author[22]{Federico Sanchez Nieto}
\author[12,4,23]{Oleksandr Tomalak}
\author[17]{Callum Wilkinson}
\author[24]{Clarence Wret}

\affil[1]{Instituto de F\'isica Corpuscular (IFIC)\\ Consejo Superior de Investigaciones Cient\'ificas (CSIC) and Universidad de Valencia (UV)\\ E-46980, Valencia, Spain}
\affil[2]{Massachusetts Institute of Technology, Cambridge, MA}
\affil[3]{Tel Aviv University, Tel Aviv, Israel}
\affil[4]{Fermi National Accelerator Laboratory, Batavia, IL 60510, USA}
\affil[5]{University College, London, London, WC1E 6BT, United Kingdom}
\affil[6]{University of Oxford, Oxford, OX1 3RH, United Kingdom}
\affil[7]{University of Pittsburgh, Pittsburgh, PA 15260, USA}
\affil[8]{University of Notre Dame, Notre Dame, IN 46556, USA}
\affil[9]{IAR/Flab, Nagoya University, Furo-cho, Chikusa-ku, Nagoya, 464-8601, Japan}
\affil[10]{Santiago de Compostela U., Instituto Galego de Fisica de Altas Enerxias, A Coru{\~n}a, Spain}
\affil[11]{Universit\"at Mainz, 55122 Mainz, Germany}
\affil[12]{University of Kentucky, Department of Physics and Astronomy, Lexington, KY 40506, USA}
\affil[13]{University of Virginia, Charlottesville, VA 22904, USA}
\affil[14]{Michigan State University, East Lansing, MI 48824, USA}
\affil[15]{University of Warwick, Coventry, CV4 7AL, United Kingdom}
\affil[16]{University of California, Berkeley, CA 94720, USA}
\affil[17]{Lawrence Berkeley National Laboratory, Berkeley, CA 94720, USA}
\affil[18]{Florida State University, Tallahassee, FL 32306, USA}
\affil[19]{University of Florida, Gainesville, FL 32611-8440, USA}
\affil[20]{Wayne State University, Detroit, MI 48202 USA}
\affil[21]{University of South Carolina, Columbia, SC 29208, USA}
\affil[22]{Universit\'e de Gen\`eve, 1211 Geneva, Switzerland}
\affil[23]{Theoretical Division, Los Alamos National Laboratory, Los Alamos, NM 87545, USA}
\affil[24]{University of Rochester, Rochester, NY 14627, USA}

\begin{document}

\rightline{FERMILAB-CONF-22-149-ND,LA-UR-21-31459}

{\let\newpage\relax\maketitle}
 
\begin{abstract}
Neutrino interaction uncertainties are a limiting factor in
current and next-generation experiments probing the fundamental
physics of neutrinos, a unique window on physics beyond the Standard
Model.  Neutrino-nucleon scattering amplitudes are an important part
of the neutrino interaction program.  
However, 
since all modern neutrino detectors are composed primarily of
heavy nuclei, knowledge of elementary neutrino-nucleon amplitudes
relies heavily on experiments performed in the 1970s and 1980s, whose statistical
and systematic precision are insufficient for current needs.  In
this white paper, we outline the motivation for attempting
measurements on hydrogen and deuterium that would improve this
knowledge, and we discuss options for making these measurements either
with the DUNE near detector or with a dedicated facility.
\end{abstract}

\tableofcontents

\section{Introduction}

Current and next generation accelerator-based neutrino experiments are
poised to answer fundamental questions about neutrinos.  Precise
neutrino scattering cross sections on target nuclei are critical to
the success of these experiments.  These cross sections are computed
using nucleon-level amplitudes  combined with nuclear models.
Regardless of whether nuclear corrections are constrained
experimentally or derived from first principles, independent knowledge
of the elementary nucleon-level amplitudes is essential.

This report explores the opportunities and challenges of a new
elementary target experiment.
Section~\ref{sec:motivation} discusses 
the need for new constraints on elementary amplitudes. 
This topic is independent of the source of new constraints
and simultaneously serves to delineate the need for, and impact of,
complementary constraints such as lattice QCD.
After surveying the current status, we 
consider the quantitative impact that improved measurements would
have for the neutrino oscillation program,
and also for physics measurements and searches
beyond neutrino interaction constraints, and beyond the Standard Model. 
Section~\ref{sec:expt} considers possible experimental configurations 
for a new H or D experiment.
Section~\ref{sec:summ} is a summary. 

\section{Scientific motivation \label{sec:motivation}}

Our understanding of nucleon-level neutrino scattering amplitudes comes
predominantly from electron scattering measurements on hydrogen and
light nuclei, and from neutrino scattering experiments. The electron
scattering measurements constrain vector-current interactions and are very precise.  Near-threshold pion electrodroduction data constrain axial-current contributions in a limited kinematic range in a model-dependent fashion~\cite{Friscic:2015tga,A1:2016emg}.
Neutrino scattering measurements constrain axial-current interactions and are less precise.  The available neutrino data on hydrogen
or light nuclei come from bubble-chamber experiments of the 1970s
and 1980s~\cite{Mann:1973pr,Barish:1977qk,Miller:1982qi,Baker:1981su,Kitagaki:1983px}.   These data have served the community well but they have
essential shortcomings.  They have poor statistical precision, coming
from the relatively low-intensity neutrino beams of an earlier era, and they have
poorly-constrained systematic uncertainties associated with
hand-scanning of events and poorly known fluxes.  Moreover, in most
cases, event-level data have been lost, and information exists only as
one-dimensional projections in publications.  Using the published bubble-chamber
data in subsequent work involves making assumptions about the details of the
analysis, such as whether or not efficiency corrections have been applied to
event rate distributions~\cite{Wilkinson:2014yfa}.

While these experiments were pioneering in their age, and probed
qualitative features of neutrino interactions that helped establish
our modern Standard Model (SM) of the strong and electroweak forces, 
they were not designed to underpin the ambitious neutrino oscillation
experiments of the current precision era.
It is clear that new and better data are needed.  However, in
order to motivate and design a new experiment, it is critical to
make this argument quantitative.
Establishing the impact of improved elementary amplitude constraints
simultaneously serves to delineate the precision targets and
potential impacts of complementary techniques such as lattice QCD. 

\subsection{Overview and status of elementary amplitudes} \label{sec2_1:amplitudes}

Given the importance of elementary amplitude constraints on the neutrino oscillation program,
it is important to explore a wide variety of complementary techniques and processes.
Any motivation for the elementary target experiment simultaneously serves as motivation
for alternative methods that can be used to obtain the same information.  Multiple determinations of the same parameter provide tests of the SM.  If discrepancies arise, something is to be learned either
about new physics or the details of the experiment.  
It is important to fully explore the scope and limitations of alternative theoretical and experimental methods in order to isolate the unique capabilities of an elementary target neutrino experiment and maximize its value.

\subsubsection{Invariant form factors} \label{sec2_1_1:invariant_ffs} 
Working at leading order in electroweak couplings, quark-level interactions with neutrinos are described by
the Lagrangian
\begin{align}
  {\cal L}_{\rm eff} &=
  - \frac{G_F}{\sqrt{2}} \left[ J^{+ \mu} J^-_{\mu} +  J^{0 \mu} J^0_\mu \right],
\end{align}
after integrating out the $W$ and $Z$ bosons. Here $G_F$ is the Fermi coupling constant, $J^\pm$ and $J^0$ are charged and neutral currents, $J^-_\mu={J^+_\mu}^\dagger$,
\begin{align}
    J^{+}_\mu &= \sum_\ell \bar{\nu}_\ell \gamma_\mu(1-\gamma_5) \ell
      + \sum_{ij} V_{ij} \bar{U}_i \gamma_\mu (1-\gamma_5) D_j ,
    \label{eq:current+} \\
    J^{0}_\mu &= \sum_f \left[ g_L^f \bar{f} \gamma_\mu (1-\gamma_5) f
      + g_R^f \bar{f} \gamma_\mu (1 + \gamma_5) f \right] ,
    \label{eq:current0}
\end{align}
and $g_{L,R}^f = I_3 (f_{L,R} ) - Q(f) \sin^2\theta_W$, with the third component of isospin $I_3$, the
electric charge in units of the proton charge $Q$, and the weak mixing angle $\theta_W$. $V_{ij}$ is the CKM matrix element relating the electroweak gauge eigenstate basis for quarks to the mass eigenstate basis.

The hadronic matrix elements $\langle p|J^+_\mu|n\rangle$, $\langle n|J^-_\mu|p\rangle$, and
$\langle N|J^0_\mu|N\rangle$ of the
currents in Equations (\ref{eq:current+}) and (\ref{eq:current0}) are decomposed into Lorentz-covariant forms of the
nucleon four-momenta, multiplied by  functions of $q^2$ known as ``form factors". For example, 
\begin{multline}
    \langle p(p') | J^{+}_{\mu} | n(p) \rangle = \bar{u}^{(p)}(p') \left\{
        \gamma_\mu F_1^\text{CC}(q^2) + \frac{i \sigma_{\mu\nu} q^\nu}{2m_N} F_2^\text{CC}(q^2) +
        \gamma_\mu \gamma_5 F_A^{CC}(q^2) 
    \right. \\ \left.
      + \frac{q_\mu}{m_N} \gamma_5 F_P^\text{CC}(q^2) \right\} u^{(n)}(p),
  \label{eq:amp}
\end{multline}
with $q^\mu = p'^\mu - p^\mu$. The corresponding expressions for the 
neutral-current scattering matrix elements $\langle p | J^{0}_{\mu} | p \rangle$ and
$\langle n|J^{0}_{\mu} | n \rangle$ involve the form factors $F_i^{\text{NC, }p}$ and $F_i^{\text{NC, }n}$.
For the vector case, these are the Dirac and Pauli form factors, $F_1(q^2)$ and $F_2(q^2)$, respectively,
which are often expressed in terms of the Sachs electric and magnetic form factors, $G_E(q^2)=F_1(q^2)+q^2
F_2(q^2)/(4m_N^2)$ and $G_M(q^2)=F_1(q^2)+F_2(q^2)$.
For the axial-vector case, one has two more form factors, $F_A(q^2)$ and $F_P(q^2)$, known as the axial
and the induced pseudoscalar form factors.  The fact that QCD respects the discrete symmetries $C$, $P$, and $T$ implies that the basis of Eq.~(\ref{eq:amp})
is complete. In the unpolarized neutrino scattering cross section, the contribution of the induced pseudoscalar form factor, $F_P(q^2)$, is suppressed by a
factor of $m_\ell^2$ (for free nucleons), so it is less
important than~$F_A(q^2)$. In polarized observables, the axial form factor contributes sizably at all energies~\cite{Pais:1971er,Bilenky:2013fra,PhysRevD.101.073002,Tomalak:2020zlv} while the pseudoscalar form factor can also be accessed either with muon neutrino beams with energies of hundreds of MeV, or with tau neutrinos~\cite{Tomalak:2020zlv}. Recent lattice calculations of the pseudoscalar form factor are contradictory between two groups~\cite{Jang:2019vkm,Park:2020axe,Jang:2020ygs,Alexandrou:2020okk}, and often give a significant difference to a conventional ansatz obtained from the assumption of partially conserved axial current (PCAC) with pion-pole dominance~\cite{Rajan:2017lxk,Alexandrou:2020okk}.
Experimental data is needed to guide theoretical calculations of the nucleon form factors.

In the following subsections, we summarize current knowledge of the 
form factors from a range of experimental and
theoretical constraints.
Constraints may be divided into three categories: form factor normalization at $q^2=0$,
form factor slopes at $q^2=0$, and general $q^2$ dependence. 
In the following text, we describe the experimental constraints on the normalization,
slope and residual shape parameters.

\subsubsection{Electromagnetic form factors} \label{sec2_1_2:electromagnetic_ffs} 
Form factor normalizations are defined by electric charges (in units of the positron charge)
and magnetic moments of
the nucleons:
\begin{align}
    G_E^N(0)= Q_N \,, \quad G_M^N(0)=\mu_N ,
\end{align}
where $Q_p=1$, $Q_n=0$, $\mu_p=2.79$, and $\mu_n=-1.91$. 
Form factor slopes are conventionally defined as charge $r_E^N$ and magnetic $r_M^N$ radii
\begin{align}
    \frac{d G_E^{N}}{dq^2}\bigg|_{q^2=0} &= \frac16 \left(r_E^N\right)^2 , \quad
    \frac{1}{G_M^N(0)} \frac{ d G_M^{N}}{dq^2}\bigg|_{q^2=0} = \frac16 \left(r_M^N\right)^2 . 
\end{align}
The most precise determination of the neutron charge radius is from low-energy neutron
scattering on the electrons of heavy nuclei, $(r_E^n)^2=-0.1161\pm0.0022~\text{fm}^2$.
For the proton charge radius, the recent development of muonic hydrogen spectroscopy
has provided the first and the most precise determination, $r_E^p=0.84087(26)(29)\, {\rm fm}$~\cite{Antognini:1900ns},
from the muonic hydrogen Lamb shift measurements. There is a $5.6\sigma$ discrepancy,
representing a $\sim 8\%$ discrepancy in the value of the slope $(r_E^p)^2$
between the muonic hydrogen result and previous determinations based on regular hydrogen spectroscopy
and electron scattering, $r_E^p=0.8751(61)\,{\rm fm}$~\cite{Mohr:2015ccw}.  
This discrepancy has become known as the proton radius puzzle and remains controversial (for reviews see Refs.~\cite{Pohl:2013yb,Carlson:2015jba,Hill:2017wzi,Peset:2021iul}).
The magnetic radii are primarily determined by electron scattering measurements,
$r_M^p = 0.776(34)(17)\,{\rm fm}$~\cite{Lee:2015jqa} and $r_M^n = 0.864^{+0.009}_{-0.008}\,{\rm fm}$~\cite{Olive:2016xmw,Belushkin:2006qa,Epstein:2014zua}.  
The general $q^2$ dependence of the vector form factors is
constrained by electron-proton scattering, and
from electron scattering on light nuclear targets, interpreted as  
electron-neutron scattering after correcting for nuclear effects.

\subsubsection{Charged-current vector form factors} \label{sec2_1_3:charged_current_ffs} 
The relevant hadronic matrix elements for charged-current processes involve 
the isovector quark current.  
Neglecting isospin violations from up- and down-quark mass terms and higher-order
electroweak effects, the isovector electroweak form factors are given
by the difference of the proton and neutron electromagnetic form factors.
Many current neutrino scattering analyses employ the
BBBA2005 parameterization~\cite{Bradford:2006yz} for the isovector
nucleon form factors.
The global data for the nucleon electromagnetic form factors
has been more recently analyzed using the $z$ expansion in Refs.~\cite{Lee:2015jqa,global}. A dedicated analysis for applications with accelerator neutrinos~\cite{Borah:2020gte}, which also provides uncertainties and correlations, has shown a 3-5\% difference for the central value of quasielastic cross sections compared to BBBA2005 input. This difference originates from the significantly different extraction of the proton magnetic form factor that was recently performed by the A1@MAMI Collaboration~\cite{Bernauer:2010wm,Bernauer:2013tpr}.

\subsubsection{Neutral-current vector form factors} \label{sec2_1_4:neutral_current_ffs} 
The neutral-current vector form factors, restricting to 3-flavor QCD,
consist of linear combinations of $u$, $d$, and $s$ quark currents, and
are thus not fully determined by the proton and neutron electromagnetic form factors.  
Many current neutrino scattering analyses neglect strange- and other
heavy-quark contributions, and assume a common dipole $Q^2$ dependence for
the remaining isoscalar and isovector combinations~\cite{Ahrens:1986xe,Andreopoulos:2015wxa}.  
It may be necessary to revisit these approximations with future precision
neutral-current neutrino data.
A discussion and further references for the vector form factor
normalization and slopes within 3-flavor QCD are found in Sec.~4.1 and
Appendix~B of Ref.~\cite{Hill:2014yxa}.  

\subsubsection{Axial form factors: charged current} \label{sec2_1_5:axial_ff_cc} 
Experimental constraints on the nucleon axial form factor may again be divided into three categories: form factor normalization at $q^2=0$,
form factor slopes at $q^2=0$, and general $q^2$ dependence. 
Form factor normalizations are determined from the neutron beta decay, 
\begin{align}
  F_A(0) &= g_A \,, 
\end{align}
with~\cite{Olive:2016xmw} $g_A=-1.2723(23)$.
The axial radius is defined analogously to the vector radii,
\begin{align}\label{eq:rA2}
  \frac{1}{F_A(0)} \frac{d F_A}{dq^2}\bigg|_{q^2=0} &= \frac16 \left(r_A\right)^2 \,. 
\end{align}
The axial radius, and general $q^2$ dependence of $F_A$, is
constrained by several processes.
Neutrino-deuteron scattering, interpreted as
neutrino-neutron scattering after correcting for nuclear effects,
provides the most direct access
to $F_A$ over a broad $q^2$ range.  A recent analysis~\cite{Meyer:2016oeg},
using the $z$ expansion, obtains an axial radius $r_A^2=0.46(22)\,{\rm fm}^2$
from existing bubble chamber data~\cite{Mann:1973pr,Barish:1977qk,Miller:1982qi,Baker:1981su,
  Kitagaki:1983px,Kitagaki:1990vs}.
 
Existing constraints on $F_A(q^2)$ inferred from charged pion
electroproduction~\cite{Amaldi:1972vf,Brauel:1973cw,
  DelGuerra:1975uiy,DelGuerra:1976uj,Esaulov:1978ed,Amaldi:1970tg,Bloom:1973fn,Joos:1976ng,Choi:1993vt,
  Liesenfeld:1999mv,Friscic:2016tbx}
have similar statistical power~\cite{Bhattacharya:2011ah} but suffer from
model-dependent corrections to the chiral limit~\cite{Bernard:2001rs}.
The muon capture process $\mu^- p \to \nu_\mu n$ from the
muonic hydrogen ground state probes a combination of
$F_A(q_0^2)$ and $F_P(q_0^2)$,
where $q_0^2=-0.88\,m_\mu^2$~\cite{Andreev:2012fj}.  Interpreted as an axial radius, the measured rate implies~\cite{Hill:2017wgb} $r_A^2=0.46(24)\,{\rm fm}^2$.

\subsubsection{Axial form factors: neutral current} \label{sec2_1_6:axial_ff_nc}
The neutral-current axial-vector form factors, restricting to 3-flavor QCD,
consist of linear combinations of $u$, $d$, and $s$ quark currents.
Many current neutrino scattering analyses account for strange- and other
heavy-quark contributions by rescaling the normalization at $q^2=0$
that would be obtained from the purely isovector case: 
\begin{align}
 F_A^{\rm NC}(0) = F_A^{\rm CC}(0)\big( 1 + \eta \big) \,,
\end{align}
with default value $\eta = 0.12$, and assuming a common dipole $Q^2$
dependence~\cite{Ahrens:1986xe,Andreopoulos:2015wxa}.
It may be necessary to revisit these approximations with future precision
neutral-current neutrino data.
A discussion and further references for the axial-vector form factor
normalization and slopes within 3-flavor QCD is found in Sec.~4.2 and
Appendix~B of Ref.~\cite{Hill:2014yxa}.  

\subsubsection{Form factor parameterizations} \label{sec2_1_7:ff_parameterizations}
A range of parameterizations has been used for the
form factors appearing in neutrino scattering analyses.
Historical benchmarks include the dipole ansatz for the axial form factor~\cite{LlewellynSmith:1971uhs},
\begin{equation}
    F_A(q^2) = \frac{g_A}{(1-q^2/M_A^2)^2},
    \label{eq:FA:dipole}
\end{equation}
and ratios of polynomials for the
vector form factors~\cite{Olsson:1978dw}.  A variety of other forms
have been used more recently~\cite{Sick:2003gm,Bodek:2007ym,Bernauer:2013tpr,Amaro:2015lga}. 
The so-called $z$ expansion provides a model-independent
description of form factor shape and quantification of the shape uncertainty. 
The formalism for the $z$ expansion and the nucleon form factors is
  described in  Refs.~\cite{Bhattacharya:2011ah,Hill:2010yb}, and
  several applications are  found in
  Refs.~\cite{Lorenz:2014vha,Epstein:2014zua,Lee:2015jqa,Bhattacharya:2015mpa}.
  Related formalism and applications may
  be found in~\cite{Hill:2006ub, Bourrely:1980gp,
    Boyd:1994tt,Boyd:1995sq,Lellouch:1995yv,Caprini:1997mu,Arnesen:2005ez,
    Becher:2005bg,Hill:2006bq,Bourrely:2008za,Bharucha:2010im,Amhis:2014hma,Bouchard:2013pna,Bailey:2015dka,Horgan:2013hoa,Lattice:2015tia,Detmold:2015aaa}.
The underlying analytic structure of
the form factor implies that a change of variable from $q^2$ to $z$,
\begin{align}
  z(q^2) &= \dfrac{ \sqrt{t_{\rm cut} - q^2} - \sqrt{t_{\rm cut} - t_0} }{
    \sqrt{t_{\rm cut} - q^2} + \sqrt{t_{\rm cut} - t_0} } \,,
\end{align}
maps the form factor shape onto a convergent Taylor expansion throughout
the entire spacelike scattering region:
\begin{align}
 F(q^2) &= \sum_{k=0}^\infty a_k [z(q^2)]^k \,, 
\end{align}
for generic form factor $F$. Here $a_k$ are dimensionless numbers encoding nucleon structure, 
$t_{\rm cut}$ is the mass
of the lightest state that can be produced by the current under consideration,
and $t_0$ is a free parameter chosen for convenience.   The number
of relevant parameters is determined a priori by the kinematic
range and precision of data.   For example, in the case of the
axial form factor, for $0<-q^2<1\,{\rm GeV}^2$, we can choose $t_0$
so that $|z|<0.23$, and it can be readily seen that quadratic,
cubic and quartic terms enter at the level of $5\%$, $1\%$ and $0.3\%$. 

\subsubsection{Quasielastic hyperon production} \label{sec:qehyperon}

Flavor-changing charged currents in the Standard Model turning light quarks into strange ones make possible the quasielastic production of hyperons in $\bar\nu$-nucleon scattering.  One has the following $\Delta S = 1$, Cabibbo ($V_{us}$) suppressed reactions
\begin{eqnarray}
\bar \nu_\ell(k) \, p(p) &\rightarrow&  \ell^+(k') \, Y^0(p') \,, \quad Y^0=\Lambda,\Sigma^0, \label{eq:anuY1} \\
\bar\nu_\ell(k) \, n(p) &\rightarrow& \ell^+(k')  \, \Sigma^-(p') \,, \label{eq:anuY2}
\end{eqnarray}
theoretically studied in Refs.~\cite{Singh:2006xp,Mintz:2007zz,Kuzmin:2008zz} following early research in the 1960s. In analogy to Eq.~\ref{eq:amp}, the matrix element of the weak current is given by 
\begin{equation}
    \langle Y(p') | J_{-}^{\alpha} | N(p) \rangle = V^\alpha - A^\alpha,
    \label{eq:CChyperon}
\end{equation}
where
\begin{eqnarray}
V^\alpha &=& \bar u_Y(p') \left[ \gamma^\alpha f_1(q^2)+ i \sigma^{\alpha\beta} \frac{q_\beta}{M+M_Y} f_2(q^2) + \frac{q^\alpha}{M_Y} f_3(q^2) \right] u(p) \,, \\
A^\alpha &=&  \bar u_Y(p') \left[\gamma^\alpha g_1(q^2) +   i \sigma^{\alpha\beta} \frac{q_\beta}{M+M_Y} g_2(q^2)  + \frac{q^\alpha}{M_Y} g_3(q^2)  \right] \gamma_5  \,u(p) \,.
\end{eqnarray}
Present knowledge about the transition form factors  $\{f_{1-3},g_{1-3}\}$ is far from satisfactory. It comes mostly from hyperon decays and is then limited to very low momentum transfers. Assuming SU(3) symmetry, these form factors can be related to the electromagnetic and axial form factors of nucleons (see for example Table II of Ref.~\cite{Singh:2006xp}). In this limit, $f_3 = g_2 =0$. SU(3) breaking corrections, which can be systematically studied using chiral perturbation theory~\cite{Zhu:2000zf} only at low momentum transfers, are not well known. The available experimental information (see Ref.~\cite{Formaggio:2013kya} and references therein) is limited by low statistics. Because of their weak non-leptonic decays, $Y \rightarrow N \, \pi$, hyperons become a source of pions in experiments with antineutrino beams at incident energies of 500-700~MeV, being potentially relevant for the short-baseline program at Fermilab. 

\subsubsection{Spin polarization physics} \label{sec2_1_8:spin_observables}

Polarization observables in neutrino-nucleus interactions could provide new constraints on 
neutrino-nucleon and neutrino-nucleus cross sections for use in the next generation of long-baseline neutrino oscillation experiments.
Measurement of the kinematic dependence of the unpolarized cross section~\cite{Bernauer:2010wm,Bernauer:2013tpr,Xiong:2019umf,Punjabi:2015bba,Ganichot:1972mb,Bosted:1989hy} is the traditional way to access the electromagnetic form factors of nucleons and light nuclei defined in Section~\ref{sec2_1_1:invariant_ffs} and discussed in Section~\ref{sec2_1_2:electromagnetic_ffs}. With the development of experimental target polarization techniques and new methods to detect the polarization of recoiling particles, complementary access to the electromagnetic structure of the nucleon~\cite{Dombey:1969wk,Akhiezer:1974em} became available a few decades ago~\cite{Perdrisat:2006hj,Jones:1999rz,Gayou:2001qd,Punjabi:2005wq,Puckett:2010ac,Ron:2011rd,Zhan:2011ji}. These days, polarized observables are as important in accessing the electromagnetic form factors as their unpolarized counterparts, 
especially 
in kinematic regions with large momentum transfers.
For example, double-polarization observables provide the best constraint on the ratio of electric and magnetic form factors of the proton at large momentum transfer.

Neutrino scattering measurements with polarized targets, as well as the detection of polarization of recoiling particles, can provide 
complementary access to the axial structure of nucleons and nuclei~\cite{ Lee:1962jm,Florescu:1968zz,Pais:1971er,Cheng:1971mx,Tarrach:1974da,Oliver:1974de,Kim:1978fx,Ridener:1984tb,Ridener:1986ey,Bilenky:2013fra,Bilenky:2013iua,Graczyk:2019xwg,Graczyk:2019opm,Fatima:2018tzs,Fatima:2018gjy,Hagiwara:2003di,Hagiwara:2004gs,Graczyk:2004vg,Graczyk:2004uy,Bourrely:2004iy,Kuzmin:2004ke,Kuzmin:2004yb,Aoki:2005wb,Aoki:2005kc,Sobczyk:2019urm,Fatima:2020pvv,Akbar:2016awk}. Contrary to the electromagnetic interactions, observables with access to the spin of one particle are not suppressed in weak interactions. The corresponding single-spin asymmetries are large and will have a comparable number of events to the unpolarized cross section measurements. 
In particular, the nucleon axial form factor, c.f. Section~\ref{sec2_1_5:axial_ff_cc}, can be extracted with GeV antineutrino beams from the single-spin asymmetry with a target polarized along the beam direction, or with a measurement of the longitudinal polarization of recoil nucleons. 
Polarization observables can also help disentangle the nucleon pseudoscalar form factors in experiments with neutrinos~\cite{Tomalak:2020zlv}. The form factors can be extracted from precise measurements of asymmetries in the scattering cross sections of muon (anti)neutrinos of hundreds of MeV energy,  or from transverse target and recoil nucleon asymmetries in the scattering cross sections of tau (anti)neutrinos above the tau production threshold $E_\nu \ge 3.5~\mathrm{GeV}$.

Hyperons produced in $\bar\nu$-nucleon interactions (see Section~\ref{sec:qehyperon}) are polarized and their polarization can be inferred from the angular distribution of the $Y \rightarrow N \, \pi$ decay products. A non-zero polarization in the direction perpendicular to the scattering plane would be a signature of T-invariance violation \cite{Fatima:2018gjy}.  
Polarization observables can also provide us access to the pion-production amplitudes~\cite{Graczyk:2017rti,Graczyk:2019blt,Graczyk:2021oyl} and axial contributions in inelastic neutrino-nucleon and neutrino-nucleus processes discussed in Section~\ref{sec2_3:inelastic_scattering} below.
Besides target and recoil polarizations, axial contributions can be accessed from the polarization of outgoing leptons. Reconstruction of the lepton spin direction from the decay products should be investigated in modern and near-future neutrino scattering experiments.

The measurement of polarized observables in neutrino interactions also provides a path toward the observation of nucleon structure that is complementary to the physics to be explored at the upcoming Electron-Ion Collider.
LBNF beam will have intensities and energies that enable lepton scattering with kinematic overlap at the lower energies produced at the EIC and can explore topics related to precision nucleon structure. In particular, certain contributions to the formulation of Generalized Parton Distributions, including certain single- and double-spin asymmetries are only accessible through polarized scattering with neutrinos. Experiments utilizing intense neutrino beams at the energies to be used at long-baseline oscillation experiments can extend EIC physics below the range of the EIC and QCD beyond the theoretical limit of perturbative QCD \cite{AbdulKhalek:2021frt}.

\subsection{Complementary constraints on elementary amplitudes} \label{sec2_2:constraints_on_amplitudes}

This section provides an overview of complementary techniques to determine the elementary amplitudes, including discussion of future prospects.  To simplify the discussion, we focus attention on techniques to constrain 
the nucleon axial form factor, $F_A(q^2)$, since this amplitude is most directly impacted by elementary target neutrino data.
See the related discussion in Ref.~\cite{Hill:2017wgb}. 

\subsubsection{Lattice QCD} \label{sec2_2_1:lattice_QCD}
 
Lattice QCD (LQCD) is a method for computing low-energy properties of hadrons based on first principles, starting from the QCD Lagrangian. This method has reached a mature state for meson properties~\cite{Aoki:2016frl}. Nucleons present an additional challenge for lattice simulations, owing to a well-known noise problem~\cite{DeGrand:1990ss}. A variety of approaches are being taken to explore and address the simultaneous challenges of excited states, lattice size, finite volume, as well as statistical noise.

Current and projected constraints are discussed in 
the USQCD white paper~\cite{Kronfeld:2019nfb}.
There has also been rapidly increasing interest in the LQCD community in calculating nucleon and nuclear structure quantities that are important to  neutrino-oscillation physics.
Most available calculations are of the nucleon isovector axial and the electromagnetic form factors.
The methodology for the calculation is well established.  Many collaborations have reported form factors calculated at the physical pion mass.  Some calculations attempt to take the continuum limit by extrapolating the results as functions of lattice spacing, and evaluating the extrapolations to a zero lattice spacing.
Improvements in the precision of current lattice results are being pursued and are now possible with the increasing computational resources available to the lattice community.
Other calculations currently being pursued include: form factors  
related to second-class currents ($F_S$ and $F_T$), which are relevant to quasielastic neutrino and antineutrino scattering;
and moments of the strange parton density functions, which will help us to understand the ``NuTeV anomaly''.

There are a number of challenging ongoing calculations that will not only require computational resources but also personpower to develop the tools and establish the methodology needed to move forward.
The transition form factors, for example, are complicated, since one needs to discern the resonances and multibody final states as we try to make realistic calculations at lighter pion mass.
The Bjorken-$x$--dependent parton densities to improve our knowledge of the large-$x$ limit, and the ``hadron tensor'' method to calculate quantities for shallow and deep inelastic scattering are being developed.
There are also ongoing efforts to expand calculations of the axial form factors in light nuclei, which provide inputs into the three-body forces in nuclear EFTs, particularly those aspects of these forces that are challenging to access experimentally.

We remark that there is a tension between axial form factor extractions from experimental deuterium bubble chamber data~\cite{Meyer:2016oeg}, and LQCD results, as shown in Fig.~3 of Ref.~\cite{Meyer:2022mix}.
The tension is strongest at large $Q^2$, where the deuterium extraction is lower than the lattice prediction. When translated to the nucleon quasielastic cross section, this discrepancy suggests that a 30-40$\%$ increase is needed to match the LQCD.  
Recent high-statistics data on nuclear targets cannot directly resolve such discrepancies due to nuclear modeling uncertainties. 
It is interesting to note that a recent MicroBooNE interaction model~\cite{MicroBooNE:2021ccs} preferred a systematic shift that enhances the quasielastic cross section contribution by as much as 20$\%$.
The consistency of the several existing LQCD results, simulated with different parameters and methods, suggests that the tension is unlikely to be due to lattice systematics.
New elementary target neutrino data would provide a critical input to resolve such discrepancies.

\subsubsection{Muon capture} \label{sec2_2_2:muon_capture}

The process $\mu^- p \to \nu_\mu n$ from the muonic hydrogen ground state is sensitive to nucleon structure, and the existing rate measurement~\cite{Andreev:2012fj} provides a competitive determination of the nucleon axial radius~\cite{Hill:2017wgb} compared to neutrino scattering data from deuterium bubble chambers~\cite{Meyer:2016oeg}.
A future improved measurement of this rate is 
feasible~\cite{Hill:2017wgb} and could reduce
the axial radius uncertainty by a factor $\sim 3$.
This would make $r_A^2$ a subdominant contribution to the error budget for $E_\nu \sim {\rm GeV}$ neutrino cross sections (taking into account residual shape uncertainty from higher order terms in the Taylor expansion of the form factor, $F_A(q^2)= F_A(0)(1+ r_A^2 q^2/6 + \dots)$, as determined from the $z$ expansion fit to deuterium bubble chamber data~\cite{Meyer:2016oeg}).  
The muon capture rate measurement also tests lepton universality, constraining a combination of $g_A$ and $r_A^2$ measured in muon systems.

\subsubsection{Parity-violating electron scattering} \label{sec2_2_3:PV_electron_scattering}

Parity-violating electron-nucleon elastic scattering~\cite{Beise:2004py,Androic:2009aa}, induced by weak $Z^0$ exchange, is a probe of the isovector quark current matrix element needed for charged current neutrino processes, but simultaneously involves also isoscalar and strange quark contributions that must be independently constrained. Available data do not have discriminating power to reliably extract axial radius or form factor shape information.

\subsubsection{Pion electroproduction} \label{sec2_2_4:pion_electroproduction}

Fits to pion electroproduction data~\cite{Friscic:2015tga,A1:2016emg} have historically contributed to the determination of the axial radius, with a small quoted uncertainty that can be traced to the assumed dipole form factor constraint. The statistical power of available data would be comparable to the neutrino-deuteron scattering determination, but the interpretation relies on extrapolations beyond the regime of low energies where chiral corrections are controlled. Pion electroproduction provides so far the only scattering data used for the extraction of the pseudoscalar form factor~\cite{Choi:1993vt}.

\subsubsection{\texorpdfstring{\boldmath $e^+d$ and $e^-d$ scattering}{e+d and e-d scattering}} \label{sec2_2_5:electron_deuteron_scattering}

The reaction $e^+ d \to \bar{\nu}_e pp$ probes the 
axial nucleon form factor.  Like the process 
$\nu_\ell d \to \ell^- pp$, there are relatively small 
nuclear corrections.   With a mono-energetic positron beam, 
kinematic measurements of the two final state protons would allow 
a complete determination of the missing neutrino energy and direction.  We are not aware of a detection of this process 
or existing proposal for future measurement. 

\subsection{Inelastic processes} \label{sec2_3:inelastic_scattering}

Inelastic scattering on nucleons encompasses a large variety of processes that take place when (anti)neutrinos interact with matter \cite{Formaggio:2013kya,Alvarez-Ruso:2014bla,Nakamura:2016cnn}. For few-GeV neutrinos, inelastic interactions account for a sizable fraction of the response and leave their imprint in the detected signal. While single pion production is the most relevant inelastic reaction \cite{Formaggio:2013kya,Alvarez-Ruso:2017oui}, nucleons, when illuminated by neutrinos, can also radiate photons~\cite{Hill:2009ek,Zhang:2012aka,Wang:2013wva} and, as the energy transferred to the target increases, can emit multiple pions \cite{Hernandez:2007ej,Nakamura:2015rta} or $\eta$ mesons \cite{Nakamura:2015rta}. Hadrons with strangeness can also be produced. Associated ($\Delta S =0$) strangeness production~\cite{Formaggio:2013kya,Nakamura:2015rta} leading to $YK$ and $Y K \pi$ final states, where $Y$ denotes hyperons, $Y=\Lambda, \Sigma, \Xi \ldots$, has the largest cross sections but also the highest thresholds. Cabibbo ($V_{us}$) suppressed (anti)neutrino-nucleon interactions lead to non zero net strangeness production ($\Delta S =\pm 1$)~\cite{RafiAlam:2010kf,Alam:2012zz,Ren:2015bsa,BenitezGalan:2021jdm}, with smaller cross sections but lower thresholds. In the $\Delta S = -1$ channel, the interesting $\Lambda(1405)$ state, extensively studied in hadron-physics experiments, can be directly excited~\cite{Ren:2015bsa} (see the right panel of Fig.~\ref{fig:mesonproduction}). The rich phenomenology of (anti)neutrino-nucleon inelastic scattering is illustrated in Table~\ref{tab:mB} where the processes with final-state meson-baryon pairs from the low-lying octets are listed. Differences in the particle composition for interactions induced by neutrinos or antineutrinos are apparent, particularly for $\Delta S =\pm 1$. Differences in the detector responses to these modes will lead to systematic uncertainties in oscillation measurements. 
\begin{table}[t]
    \centering
       \caption{(Anti)neutrino-nucleon scattering with meson-baryon pairs from the low-lying octets in the final state. In each row, hadron pairs are listed in order of increasing production threshold. Notice that processes with $\Delta S =-1(+1)$ induced by $\nu(\bar\nu$) as well as NC ones with both $\Delta S = \pm 1$ are suppressed in the Standard Model and, therefore, very rare.}
    \label{tab:mB}
    \vspace{2mm}
    \begin{tabular}{|c|c|c|}
    \hline
    CC ($W^-$) & CC ($W^+$) &  NC ($Z$) \\
    \hline
    $\nu \, p \rightarrow \ell^- \, X^{++}$ & $\bar\nu \, p \rightarrow \ell^+ \, X^{0}$ &  $\nu(\bar\nu) \, p \rightarrow \nu(\bar\nu) \, X^{+}$ \\ 
    $\nu \, n \rightarrow \ell^- \, X^{+}$ & $\bar\nu \, n \rightarrow \ell^+ \, X^{-}$ &  $\nu(\bar\nu) \, n \rightarrow \nu(\bar\nu) \, X^{0}$ \\
    \hline
    \end{tabular}
    \\
    \begin{tabular}{|c|l|}
    \hline
    \multirow{4}{*}{$\Delta S = 0$} 
    & $X^{++}=$ $p \pi^+$, $\Sigma^+ K^+$\\
    & $X^{+}=$ $p \pi^0$, $n \pi^+$, $p \eta$, $\Lambda K^+$, $\Sigma^0 K^+$, $\Sigma^+ K^0$\\
    & $X^{0}=$ $n \pi^0$, $p \pi^-$, $n \eta$, $\Lambda K^0$, $\Sigma^0 K^0$, $\Sigma^- K^+$\\
    & $X^{-}=$ $n \pi^-$, $\Sigma^- K^0$ \\
    \hline
    \multirow{4}{*}{$\Delta S = -1$} 
    & $X^{++}=$ $\Sigma^+ \pi^+$\\
    & $X^{+}=$ $\Lambda \pi^+$, $\Sigma^+ \pi^0$, $\Sigma^0 \pi^+$, $p \bar K^0$, $\Sigma^+ \eta$, $\Xi^0 K^+$ \\
    & $X^{0}=$  $\Lambda \pi^0$, $\Sigma^0 \pi^0$, $\Sigma^+ \pi^-$, $\Sigma^- \pi^+$, $p K^-$, $n \bar K^0$, $\Lambda \eta$, $\Sigma^0 \eta$, $\Xi^0 K^0$,  $\Xi^- K^+$ \\
    & $X^{-}=$ $\Lambda \pi^-$, $\Sigma^0 \pi^-$, $\Sigma^- \pi^0$, $n K^-$, $\Sigma^- \eta$,  $\Xi^- K^0$\\
    \hline
    \multirow{3}{*}{$\Delta S = +1$} 
    & $X^{++}=$ $p K^+$ \\
    & $X^{+}=$ $p K^0$, $n K^+$ \\
    & $X^{0}=$ $n K^0$ \\
    \hline
    \end{tabular}
\end{table}

\begin{figure}[t]
    \centering
    \includegraphics[width=0.85\textwidth]{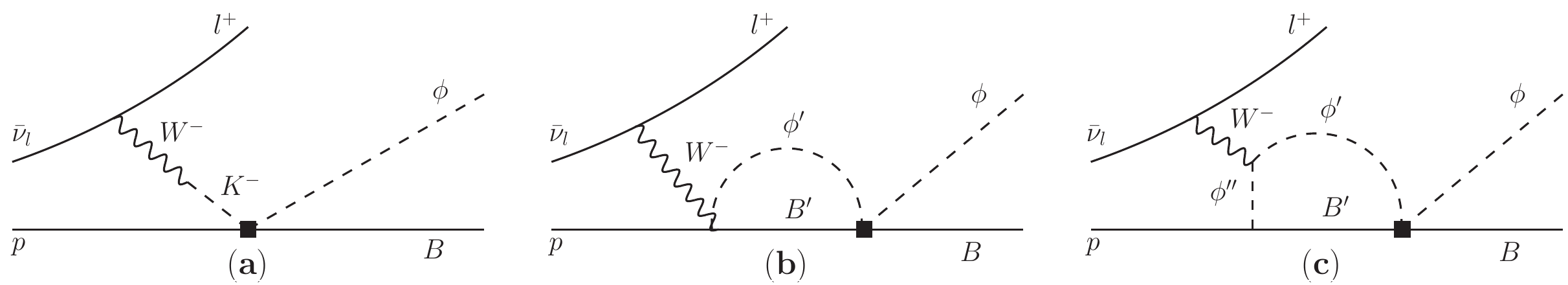}
    \caption{Diagrams contributing to $\Delta S=-1$ meson-baryon production in the chiral unitary model of Ref.~\cite{Ren:2015bsa}. The solid square represents the strong meson-baryon $T$-matrix in coupled channels.}
    \label{fig:weakuch}
\end{figure}

\begin{figure}[t]
    \centering
    \includegraphics[width=0.43\textwidth]{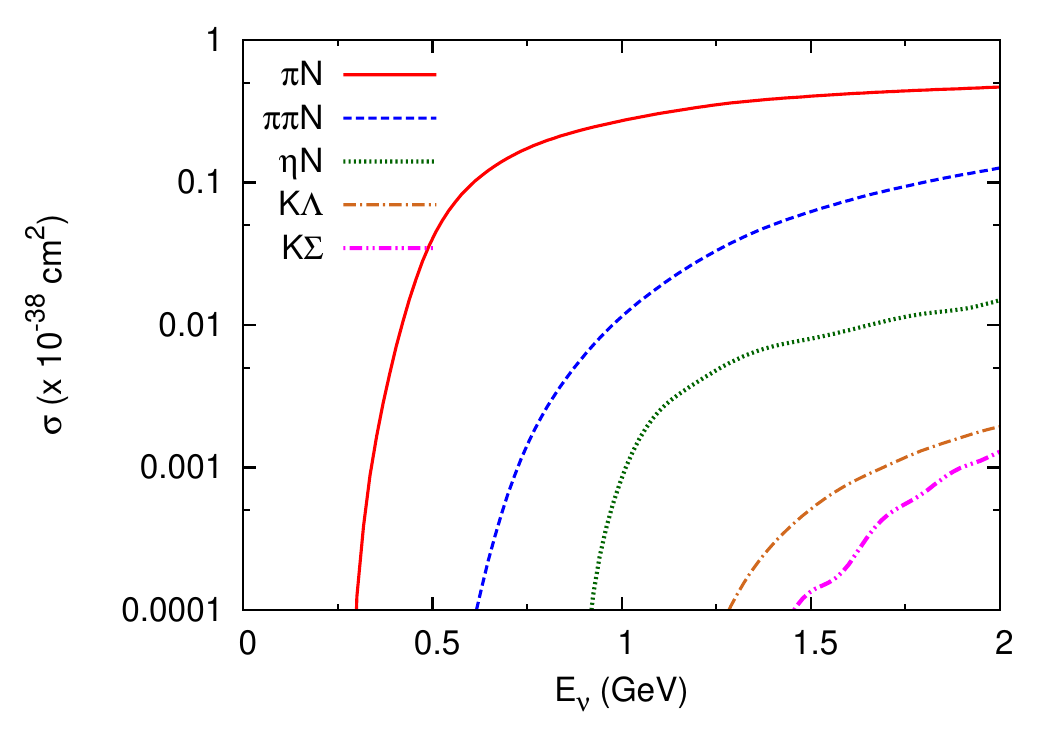}
    \includegraphics[width=0.42\textwidth]{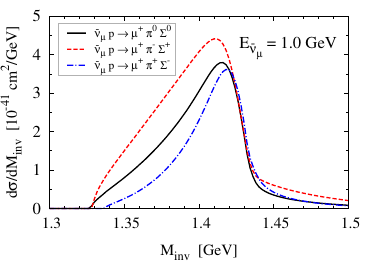}
    \caption{Left panel: cross section for different CC $\nu_\mu n$ final states as predicted by the DCC model~\cite{Nakamura:2015rta}. Right panel: line shape of the $\Lambda(1405)$ as seen $\pi\Sigma$ production in $\bar\nu_\mu p$ scattering~\cite{Ren:2015bsa}.}
    \label{fig:mesonproduction}
\end{figure}

Close to threshold, the interaction amplitudes are determined by the approximate chiral symmetry of QCD. In this kinematic region, a systematic treatment of quantum corrections is possible using chiral perturbation theory. This has been done for single pion production in Refs.~\cite{Yao:2018pzc,Yao:2019avf} up to next-to-next-to-leading order using the covariant formulation of the theory. In this approach, amplitudes are given in terms of so called low energy constants (LECs). Many of the relevant LECs have been determined from other process and, in particular, using threshold pion photo and electroproduction data~\cite{GuerreroNavarro:2020kwb}. Remaining LECs could be fixed using neutrino-nucleon data with full kinematic reconstruction and high enough statistics in the region where the perturbative treatment is valid. LEC determination is interesting by itself but would open the possibility to use pion production as a standard candle for neutrino-flux determination with controlled theoretical errors~\cite{Duyang:2019prb}. On the other hand, this formalism is applicable close to threshold; a theoretical description covering the whole kinematics available with few-GeV neutrinos, as required for neutrino event generators, demands  phenomenological modeling. The validity of chiral amplitudes towards high $Q^2$ is often extended via phenomenological form factors (see  Refs.~\cite{Hernandez:2007qq,RafiAlam:2010kf, Alam:2012zz} for single pion and kaon production). Unitarization becomes important as the invariant mass of the final hadronic system  ($W$) increases. For neutrino interactions, unitarization in coupled channels has only been implemented by the dynamical coupled-channel (DCC) model~\cite{Nakamura:2015rta,Nakamura:2016cnn} and in the strangeness $\Delta S=-1$ sector within a chiral unitary approach~\cite{Ren:2015bsa}, diagrammatically represented in Fig.~\ref{fig:weakuch}. Selected predictions from these studies are shown in Fig.~\ref{fig:mesonproduction}.

For hadronic invariant mass $W \lesssim 2$~GeV  most of these reactions proceed through the excitation of baryon resonances. The best known example is single-pion production, which is dominated by the $\Delta(1232)$. However, even around the $\Delta(1232)$ peak, the cross section arises from the interplay of resonant and non-resonant amplitudes \cite{Hernandez:2007qq}. This interplay becomes highly non-trivial at higher invariant masses with several resonances overlapping and coupled channels. This is the shallow inelastic scattering region, where a large fraction of events at DUNE will be located. The intricacies of these dynamics and the properties of baryon resonances have been investigated in detail in partial wave analyses of large data sets available for photon, electron and pion-nucleon interactions. This information is valuable to constrain weak inelastic processes and has extensively been used in their modeling. Indeed, isospin symmetry relates the electromagnetic current to the weak vector current, while in the soft pion limit, the axial current at $Q^2 = 0$ can be expressed in terms of the pion-nucleon scattering amplitude. However, the properties of the axial current at finite $Q^2$ remain largely unknown and experimentally unconstrained. In fact, theoretical predictions like those in Fig.~\ref{fig:mesonproduction} rely on little more than educated guesses for this dependence. Lattice QCD may be able to partially fill this gap of knowledge by providing, for example, $N-\Delta(1232)$ transition form factors~\cite{Alexandrou:2010uk} at nearly physical light-quark masses but a comprehensive description of the inelastic scattering dynamics is likely out of reach.

Models of neutrino scattering in the kinematic region where the transition from the resonance to the deep-inelastic scattering regime takes place are also highly uncertain.  Achieving a realistic description of these scatters in neutrino event generators is a major challenge with direct repercussions for oscillation studies with accelerator or atmospheric neutrinos \cite{Alvarez-Ruso:2017oui,SajjadAthar:2020nvy}. The conjecture of quark-hadron duality on one hand and high-twist effects on the other are valuable tools to achieve this goal, but progress in their development is hindered by the lack of experimental nuclear-effect-free information from elementary targets. 

\begin{figure}[t]
    \centering
    \includegraphics[width=0.5\textwidth]{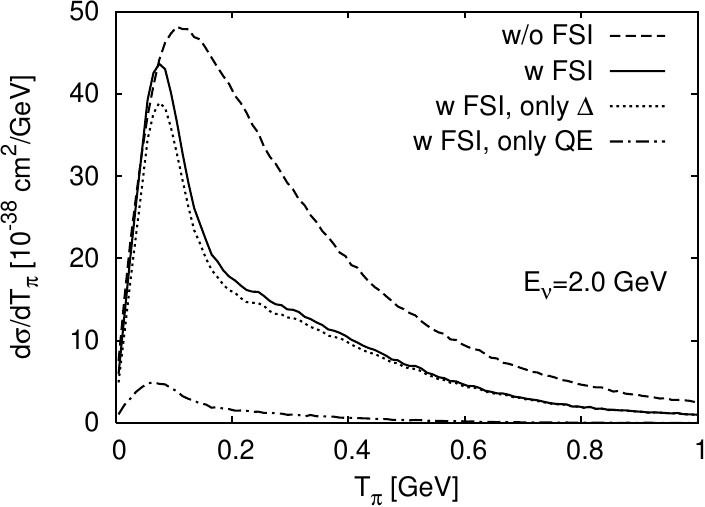}
    \caption{$\pi^+$ kinetic energy spectra for 2~GeV $ \nu_\mu$ scattering on $^{56}$Fe predicted by the GiBUU transport model~\cite{Leitner:2006ww}. The strong distortion caused by final-state interactions is apparent in the comparison of solid and dashed lines.}
    \label{fig:pifsi}
\end{figure}

Only a few of the reactions mentioned above have been experimentally measured in the past using bubble chambers with poor statistics and large uncertainties associated, in particular, with poor knowledge of the incident neutrino fluxes~\cite{Formaggio:2013kya}. Single pion production on nuclear targets has also been investigated in modern experiments such as MiniBooNE, T2K and MINERvA \cite{Mahn:2018mai,MINERvA:2021csy}. While these measurements are valuable, they hardly allow to constrain elementary amplitudes due to the distortion caused by the nuclear medium and final-state interactions \cite{Leitner:2006ww,Mosel:2019vhx} (see Fig.~\ref{fig:pifsi}).
Not only are these processes interesting by themselves as a source of currently unknown information about the axial structure of hadrons, they are relevant for the future of neutrino oscillation measurements and proton-decay searches. Their experimental study will provide solid input for the simulation of neutrino-nucleus interactions at DUNE and other facilities~\cite{Alvarez-Ruso:2017oui}. Furthermore, as explained below in Sec.~\ref{sec2_6_1:CKM_and_beta_decay}, accurate measurements of (anti)neutrino-nucleon inclusive cross sections in the non-perturbative regime, as well as its individual components, will help constrain electroweak radiative corrections to $\beta$ decay processes and lead to a more precise determination of the CKM matrix element $V_{ud}$.
Last but not least, these measurements, combined with lattice-QCD results, could reveal the presence of non-standard neutrino interactions.

\subsection{Generator tuning}
\label{sec:generatortuning}
The majority of neutrino interaction generators  factorize the interaction of a neutrino on a nuclear target into a description of the initial-state nucleon in the nucleus, and the neutrino interacting with the nucleon in the nucleon rest frame.  Hence the neutrino-nucleon interaction forms the backbone of many generators and models, whose cross section can only be measured on very light targets such as $\text{H}_2$ and $\text{D}_2$. In this way, all the issues previously discussed regarding poorly understood elementary amplitudes have direct implication on the neutrino interaction generator community.

Bubble chamber data on light targets ($\text{H}_2$, $\text{D}_2$) in the ranges relevant for GeV-scale neutrino experiments come primarily from the ANL and BNL experiments. They provided cross sections and rate measurements of CCQE, CC and NC$1\pi$, and multi-pion final states. These have been extensively used in the generator community, by NuWro~\cite{Graczyk:2014dpa,Graczyk:2009qm,Graczyk:2007bc}, GENIE~\cite{Andreopoulos:2009rq,Wilkinson:2014yfa,Rodrigues:2016xjj,Tena-Vidal:2021rpu}, NEUT~\cite{T2K:2021xwb} and GiBUU collaborators~\cite{Mosel:2015tja,Mosel:2017nzk}. Furthermore, more sophisticated interaction models currently being implemented in neutrino generators~\cite{Kabirnezhad:2017jmf,Sobczyk:2018ghy} use the neutrino-nucleon interaction as a benchmark to tune and validate against. To demonstrate the utility and longevity of such data, the single pion production measurement published by ANL in 1981~\cite{Radecky:1981fn} has 381 citations, 194 of them coming from publications written in 2009 and later. The equivalent result from BNL in 1986~\cite{Kitagaki:1986ct} has 289 citations, with 161 of them coming from 2009 or later. 

However, despite the longevity of the data, it is difficult to gain detailed insight of the neutrino-nucleon interaction because the data suffers from small sample sizes, and analysis decisions are not always well understood. As a result, the data must be carefully chosen and evaluated for analysis. For instance, it is unclear what corrections were applied in the rate measurements from ANL and BNL, and there has been a long-standing discrepancy in the single pion production results from ANL and BNL~\cite{Graczyk:2009qm, Wilkinson:2014yfa, Rodrigues:2016xjj}, possibly related to an optimistic neutrino flux model at the time. Importantly, this leads to a large uncertainty ($\sim10\%$) on the cross section of this process, which contributes as much as 50\% of the events at DUNE. Separately, the same issue has been discussed~\cite{Belusevic:1987rn, Belusevic:1988ab} as a limiting factor in methods to extract neutrino flux information in situ. Furthermore, most measurements from ANL and BNL do not provide much detail on the kinematics of the particles produced in the interactions, focusing on one-dimensional distributions in neutrino energy or invariant momentum transfer $Q^2$. In the single and multi-pion final state distributions, many selections give order $\mathcal{O}$(100) events, and are dominated by statistical uncertainties. The situation has barely improved since these ANL and BNL experiments as there has not been any new light-target experiment in the relevant energy range.

The lack of reliable high-statistics data for neutrino-nucleon scattering is a significant hindrance to understanding nuclear effects. Large uncertainties in the neutrino-nucleon interaction makes interpreting discrepancies in the nuclear environment highly non-trivial, as any effect observed on a nuclear target may come from a poor understanding of nucleon or nuclear physics. A high-statistics experiment with carefully studied systematics, as the one proposed here, can resolve long standing tensions in existing data, improve our understanding of the nucleon \emph{and} nuclear physics, and ultimately lead to significantly reduced uncertainties in future experiments, such as DUNE. 

\subsection{Impact on the oscillation program} \label{sec2_4:impact_on_oscillation_physics}

Neutrino interactions using conventional accelerator beams on nuclear
targets involve uncertainties from the neutrino flux,
elementary amplitudes and nuclear corrections.  
Independent sources of information are required in order to separate their effects. 
Accurate determination of the elementary amplitudes is a critical component needed to
break the ambiguity.  In a situation
where an accurate and complete nuclear model is fully determined
by nucleon-level inputs, it would be a straightforward task to
propagate uncertainties from inputs to observables.  In practice,
neutrino interactions are described by an iterative process, whereby
nuclear models are improved to fit available data under assumed
inputs, subject to flux and detector uncertainties.  A critical
discussion of these simultaneous uncertainties is necessary to obtain
quantitative benchmarks for future elementary target constraints.
More broadly, precision constraints on nucleon-level inputs is a
necessary step in the validation of ab initio calculations and
phenomenological models for nuclear matrix elements. 

This discussion is important beyond the
immediate goal of quantifying the impact of a new elementary target
experiment: it quantifies the current level of
uncertainty on neutrino oscillation observables owing to neutrino interactions;
and it provides goals for the precision of complementary constraints on elementary
amplitudes. 

\vspace{5mm}

\subsubsection{Flux determination} \label{sec2_4_1:fluxes}
The neutrino flux can be predicted from a GEANT4 simulation of the beamline, the target, the focusing horns, and the decay pipe~\cite{Abi:2020evt}.  Uncertainties in these predictions come primarily from hadron production modeling uncertainties, but they also contain a smaller component from the modeling of the focusing.  The total uncertainty in the flux prediction is of order 10\% and it varies smoothly with neutrino energy. Further constraining this flux uncertainty with a near detector is an important part of the program for extracting precision oscillation measurements with DUNE.

The primary purpose of the near detector at LBNF is to sample the unoscillated beam of neutrinos a short distance downstream of the target and decay pipe, in order to constrain systematic uncertainties which can then be extrapolated to the far detector, where the oscillated flux is sampled. 
However, the event counts measured by the DUNE near detector are proportional to both the flux and the cross section for the process under study.  These event counts are also affected by detector and analysis acceptance and efficiency factors, and the reconstructed neutrino energies will have biases and resolutions that come from nuclear effects, such as the amount of missing energy in neutrons, as well as detector effects. Much of the work that goes into designing the near detector is focused on trying to break the degeneracy between different components of the model.

An example of an innovation that addresses the ambiguity between uncertainties in the flux and the cross section is the PRISM technique~\cite{DUNE:2021tad}. Two of the three near detector components are designed to move as much as $34\,$m off axis.  As long as the dependence of the energy spectrum of the neutrino flux as a function of the off-axis position can be modeled, then linear combinations of the measured event rates in ND-LAr can be used to predict the oscillated spectrum of neutrinos at the far site as a function of the oscillation parameters. This extrapolation includes the effects of imperfect neutrino energy reconstruction, as long as these effects are the same for the near and far detectors.  Corrections for differences in detector acceptance, mostly due to the small size of the near detector and its reduced ability to contain events, as well as detector technology, must be accounted for in simulations.

A complementary way to break the degeneracy between the flux and the cross section is to try to isolate a sample with a well understood cross section with which the flux can be measured. 
The near detector at LBNF can make a strong absolute measurement of the inclusive flux rate via elastic $\nu-e$ scattering.  This technique has been employed by MINERvA~\cite{MINERvA:2019hhc}.  It takes advantage of the fact that the cross section for elastic scattering of neutrinos and electrons is well understood from first principles (see Ref.~\cite{Tomalak:2019ibg} and references therein).  It reduced the uncertainty on the NuMI flux from 7.8\% to 3.9\%, with the dominant uncertainty being statistical.  A study of this technique using ND-LAr is given in Ref.~\cite{Marshall:2019vdy}. Unfortunately, the energy spectrum of the neutrino beam cannot be well constrained by this method due to the intrinsic divergence of the neutrino beam, only the total rate.  The high-energy part of the flux can be measured with inverse muon decay~\cite{MINERvA:2021dhf}.  This process uses the charged-current interaction of $\nu_\mu$ on electrons, but it has a threshold of 11~GeV on the incident neutrino energy before it proceeds.

Another approach to determining the energy spectrum of the neutrino flux which is often discussed is the ``low-$\nu$'' method~\cite{Belusevic:1988ab,Bodek:2012uu}.  This method takes advantage of the fact that the neutrino cross section on a nucleus is independent of the neutrino energy in the limit of zero energy transfer to the nucleus.  In this limit, the observed event rate is proportional to the neutrino flux. However, in practice, the dynamics of the nucleus and detector limitations mean that relatively large, model-dependent, corrections are required to account for the energy dependence of small energy transfer samples that can be isolated experimentally.

A hydrogen/deuterium neutrino target can provide a separate measurement of the energy spectrum of the neutrino flux.  As long as the cross section for quasielastic neutrino scattering and any deuterium corrections are well understood (or determined self-consistently by the data), the flux can be constrained~\cite{Duyang:2019prb}.  If the detector is hermetic enough to be able to tell on an event-by-event basis that that no additional charged particles or $\pi^0$s or photons are present, then transverse-momentum balance can make even ${\bar\nu}_\mu p \rightarrow \mu^+ n$ interactions a viable channel for constraining the antineutrino spectrum in the beam.
A sample of quasielastic events can be obtained for both the neutrino and the antineutrino beam as long as deuterium is the target.  Separation of $\nu_e$ from $\nu_\mu$ and their antineutrinos which involves muon and electron ID in the detector as well as sign selection, allows the separate measurement of each component of the neutrino beam.

Constraining the shapes of the four neutrino energy spectra -- $\nu_\mu$, ${\bar\nu}_\mu$, $\nu_e$ and ${\bar\nu}_e$ for both forward horn current and reverse horn current running will reduce the systematic uncertainties in energy-dependent differential cross section measurements on the other elements in the near detector: primarily argon, carbon and iron.  Other materials, such as lead, may be used in the ND-GAr calorimeter.

Placing a hydrogen/deuterium target and detector in the Booster Neutrino Beam can also help reduce flux uncertainties in Fermilab's short baseline neutrino program.

\subsection{Contributions to a broader beyond the Standard Model program at the LBNF  \label{sec2_5:impact_on_BSM} } 

Searches for physics beyond the SM (BSM) are an increasingly important component of the long baseline neutrino program. While some new physics models (e.g., non-standard interactions) are best probed by leveraging the far detector (FD) (see e.g.\ \cite{Schwetz:2020xra}), it is typically the DUNE near detector (ND) that offers the most promising sensitivity. This is because it is closest to the meson production point and so it experiences both an intense neutrino beam and a large flux of BSM particles if they are produced in meson decays.

Comparing the deuterium bubble chamber setup proposed in this white paper to DUNE is an inherently model-dependent exercise. Considerations include geometry, background rates, detection thresholds and efficiencies, and signal yields. For this reason, we will organize our discussion around a few broad model classifications: 
\begin{enumerate}
	\item  Modified neutrino cross sections.
	\item  Prompt meson and beam-stop progenitors. 
	\item  Long-lived charged meson progenitors (and muons). 
	\item  Neutrino-induced BSM particle production. 
\end{enumerate} 
In the latter three classes of models, BSM particles are produced outside of the fiducial volume of the detector, while ultimately leaving detectable signatures within it. Examples include heavy neutral leptons, axions and axion-like particles, millicharged particles,  and dark photons \cite{Magill:2018tbb,DeRomeri:2019kic,Berryman:2019dme,Ballett:2019bgd,Harnik:2019zee,ArgoNeuT:2019ckq,Kelly:2020dda,Brdar:2020dpr}. The detector may observe scattering and/or decay signatures. The fourth scenario involves BSM particle production inside the detector, which leads to correlated signatures within the detector. For example, a neutrino may produce a BSM particle via a hard scatter with a nucleus/nucleon, producing visible hadronic activity. The exotic particle may then propagate some distance before decaying visibly. This topology mimics a ``double bang'' and has been proposed in the context of heavy neutral leptons in Super Kamiokande, IceCube, and DUNE \cite{Coloma:2017ppo,Atkinson:2021rnp}. It is also possible to search for deviations from predicted SM cross sections, provided that the SM background is well understood. In neutrino experiments with nuclear targets, this generically selects only $\nu e \rightarrow \nu e$ elastic scattering as a viable process; unfortunately this cross section is roughly four orders of magnitude smaller than CCQE on nucleons and so tends to be statistically limited (e.g., a statistical floor of $\sim 1\%$ at DUNE assuming 10,000 $\nu e$ scattering events). Neutrino-nucleon scattering offers a much lower statistical floor owing to its much larger cross section (generically enhanced by $m_N/m_e \sim 2000$), however this large event rate cannot be leveraged with nuclear targets because the underlying cross section has large theory uncertainties. Neutrino-nucleon scattering on an elementary target offers a compromise between theoretical and statistical errors that could offer improved sensitivity to new physics over neutrino electron scattering and, as we discuss below, probe models with modified neutrino-quark couplings. In addition to better theoretical control over the SM cross section predictions, the \emph{exact} kinematics of a nucleus at rest can be leveraged to limit the impact of flux determination.  We return to this point in Section~\ref{exact-kins}. 

Unlike long-baseline neutrino oscillation searches, where the ND complex assists the FD by constraining the LBNF flux, the DUNE ND complex must operate as a standalone BSM physics facility. One serious disadvantage of a LArTPC (or GArTPC) is the complicated nuclear structure of $^{40}{\rm Ar}$. Therefore, any internally-produced BSM signature suffers from large theory uncertainties on the production cross section that are, at present, poorly understood~\cite{NuSTEC:2017hzk}. Even for externally-produced BSM particles, which may for example depend only on meson production modeling and perturbatively calculable decay rates and distributions, background event rates within the DUNE ND complex are highly uncertain and can set a detection sensitivity floor. This sets a fundamental floor on the sensitivity coming from systematic uncertainties that are sometimes ignored in the BSM literature. 

A deuterium bubble chamber detector circumvents these theory uncertainty issues and so, for certain BSM signatures, may outperform the conventional LArTPC and TMS/GArTPC ND complex detectors. Nuclear effects in deuterium can be modeled using a combination of chiral perturbation theory and  ab initio methods \cite{Singh:1971md,Beane:2004ra,Acharya:2019fij}.
Uncertainties on neutrino-nucleon cross sections are sizeable  due to uncertainties on the axial form factor~\cite{Borah:2020gte}; the relative uncertainties depend on $Q^2$ and they currently range from $\pm 1\%$ to $\pm 10\%$. 
In the absence of BSM contributions, such uncertainties will be addressed by new elementary target data.  
Lattice QCD~\cite{Meyer:2022mix} 
would allow the degeneracy between QCD effects 
and potential BSM effects to be lifted.
Alternatively, if polarized observables are available, the axial form factor could be determined using unpolarized cross section data (its determination would then be statistically limited) and then used to predict polarized observables. In any case, these nucleon-level uncertainties will always be smaller than those on nuclear targets where the nucleon-level form factors are compounded by nuclear effects including meson exchange currents, final-state interactions, and initial-state nucleon momentum distributions \cite{Mosel:2016cwa}.

In what follows, we discuss each of the model classifications separately, highlighting potentially impactful physics cases that fall into each category.  Some BSM scenarios straddle multiple categories, in which case different production mechanisms and detection strategies are considered separately. We will avoid model-dependent details whenever possible, focusing on a comparison of events yields, theory uncertainties and backgrounds between hydrogen-based detectors and the DUNE ND complex. We also compare and contrast a dedicated hydrogen/deuterium bubble chamber with hydrocarbon detectors that provide hydrogen measurements by subtracting carbon target data from hydrocarbon data, a technique enhanced by using transverse kinematic imbalance (TKI) to further separate neutrino interactons on H from those on C.

As we will discuss below, event rates will be comparable in ND-LAr and ND-GAr for decaying particles, and roughly an order of magnitude larger in the LArTPC for scattering signals, assuming the scattering cross sections per nucleon are the same in the two detectors. These comparisons are expected to be true for many different production mechanisms but may vary if, for instance, coherent scattering dominates in which case the cross section on argon will be enhanced. Given comparably sized data sets, the primary drivers of physics performance will be a combination of detector performance (e.g., PID, thresholds,  efficiencies, and background mitigation) and theory uncertainties. These questions depend on the BSM signal predicted by a given model, however it is relatively easy to see that any channel with sizeable backgrounds from SM neutrino scattering must contend with irreducible theory uncertainties and that these are smaller for neutrino-nucleon scattering as compared to neutrino nucleus scattering.

\subsubsection{Modifications to neutino-nucleon cross sections. } 
\label{sec:bsm:ccqemodifications}

Models of BSM physics may be tested by searching for deviations from expected SM cross sections. This strategy has been pursued in the context of $\nu e\rightarrow \nu e$ scattering. This channel is particularly beneficial for light mediators that receive an enhancement due to the low momentum transfers involved i.e.\ $\dd \sigma/ \dd Q^2 \sim 1/Q^4$, in contrast to the suppression that characterizes weak interaction cross sections $ \dd \sigma /\dd Q^2 \sim G_F^2$. The neutrino-electron elastic scattering cross section is calculable at the per-mille level of precision \cite{Tomalak:2019ibg}, and one can therefore conduct statistically-limited studies. Applications of DUNE $\nu e$ scattering data include constraining the weak mixing angle \cite{deGouvea:2019wav}, measuring neutrino electromagnetic properties \cite{Mathur:2021trm}, and searching for millicharged particles \cite{Magill:2018tbb,Harnik:2019zee,ArgoNeuT:2019ckq}. Similar strategies have been pursued using coherent elastic neutrino-nucleus scattering which also has a SM prediction at the per-mille level \cite{Tomalak:2020zfh}.

Models in which neutrinos couple directly to nucleons via new forces are, by way of contrast, very difficult to constrain far below the level of weak cross sections. The reason is simple: SM predictions for neutrino-nucleus cross sections carry sizable $\sim 30\%$ theoretical uncertainties. Neutrino electron-scattering can be probed with the DUNE ND LArTPC at the $\sim 1\%$ level \cite{deGouvea:2019wav}, and is statistically limited. Since neutrino-nucleon cross sections are three to four orders of magnitude larger than neutrino-electron cross sections any search strategy relying on neutrino-nucleus scattering is guaranteed to be limited by theoretical uncertainties.  
The ultimate BSM impact of a hydrogen-rich sample of scattering events will then be determined by the extent to which the neutrino nucleon cross section can be predicted, either from first principles or by data-driven means. 

Neglecting BSM contributions, the axial form factor can be extracted from data (more generally, constraints on BSM parameters can be determined self-consistently together with the axial form factor). 
For example, it is well known that the spin-averaged CCQE differential cross section can be written using the ``ABC formula'', \cite{LlewellynSmith:1971uhs,Formaggio:2012cpf} for $\pbar{\nu}_\ell N \rightarrow \ell^\pm  N'$
\begin{equation}
    \dv{\sigma_{\nu/\bar{\nu}}}{Q^2} = \frac{G_F^2 |V_{ud}|^2 M}{E_\nu^2} \times \qty[ A \pm B \qty(\frac{s-u}{M^2}) + C \qty(\frac{s-u}{M^2})^2] ~,
\end{equation}
where $A$, $B$ and $C$ depend on the hadronic form factors. The positive sign in the $B$ term corresponds to the neutrino cross section, and the negative sign corresponds to the antineutrino cross section.  The factor $B$ has a particularly simple expression: $B= Q^2/M^2 \times F_A (F_1+F_2)$. Assuming fixed vector form factors, one can extact $F_A$ from the $\bar{\nu}$-$\nu$ cross section difference, 
\begin{equation}
    \dv{\sigma_{\nu}}{Q^2}- \dv{\sigma_{\bar\nu}}{Q^2}= \frac{G_F^2 |V_{ud}|^2 }{E_\nu^2 M} \times 2(s-u) \times \frac{Q^2}{M^2} F_A(Q^2) \qty[F_1(Q^2)+F_2(Q^2)]~.
\end{equation}
Given $\sim 1\%$ level extractions of the vector form factors as presented in \cite{Borah:2020gte}, and a modern treatment of radiative corrections~\cite{Hill:2016gdf,Tomalak:2021qrg}, one could realistically expect a $\sim 1\%$ level determination of the axial form factor assuming that experimental systematics and statistics are under control. Depending on the progress of lattice QCD, an experimental extraction of $F_A(Q^2)$ could serve as a test of the SM on its own since lattice QCD predictions of $F_A(Q^2)$ are made with no adjustable parameters. 

All remaining linearly-independent experimental observables, such as the summed CCQE cross sections $\dd \sigma_{\nu} + \dd \sigma_{\bar{\nu}}$, neutral-current scattering and other target or polarized lepton distributions~\cite{Tomalak:2020zlv}, may then be predicted\footnote{In principle the pseudoscalar form factor should also be extracted from data, however it is explicitly suppressed by a small factor of $m_\ell^2/Q^2 \ll1$ in regions of large momentum transfer such that larger relative uncertainties can be tolerated. An estimate, accurate to around $\sim 10\%$, can be obtained using a partially conserved axial current (PCAC) ansatz and the extracted axial form factor.}  given $F_A(Q^2)$. 

Models in the BSM landscape that benefit from precise neutrino-nucleon cross sections include new neutrino-hadron interactions, such as those in Refs.~\cite{Pospelov:2011ha,Batell:2014yra,Farzan:2018gtr}. Using CCQE as a signal channel, one could probe flavor-dependent structures. This idea is discussed within the context of non-standard interactions and SM effective field theory (SMEFT) in~\cite{Falkowski:2018dmy,Falkowski:2021bkq}. Projections in~\cite{Falkowski:2018dmy} rely on charged-current to neutral-current cross section ratios $\sigma_{\rm CC} /\sigma_{\rm NC}$, and explicitly neglect uncontrolled theoretical uncertainties related to argon's complex nuclear structure and non-zero isospin. These issues could be evaded by using nucleon targets whose form factors are well understood and constrained. A simple application would be the extraction of left- and right-handed couplings to quarks which could then be compared to precise SM determinations~\cite{Hill:2019xqk}. Deviations from the SM can be interpreted as stemming from higher-dimensional SMEFT operators \cite{Falkowski:2018dmy}.

\subsubsection{Prompt meson decay and progenitors in the target} 
\label{sec:bsm:promptprogenitors}

This category refers to BSM particles that are produced primarily in the target either through proton bremsstrahlung, prompt meson decay, hard inelastic proton scattering or other mechanisms that are displaced no more than a few meters from the target~\cite{deNiverville:2011it}. These exotic particles are necessarily neutral and weakly interacting, and they may penetrate through the intervening dirt on the way to the detector. The produced flux of BSM particles is unfocused and is diluted by geometric effects scaling as $\sim 1/d^2$ with $d$ the distance between the target and the detector. The ND complex will be 574~m from the target, while the proposed bubble chamber hall will lie somewhere between the muon alcove and the DUNE ND hall; for concreteness we will assume the bubble chamber hall will be placed 200~m upstream of the DUNE ND complex i.e., 374~m from the target.  Therefore, the flux of any BSM particles produced in the target is $\sim (574/374)^2\approx 2.4$ times as intense at the bubble chamber hall as compared to the DUNE ND complex. This prediction is essentially independent of any BSM modeling details, being driven exclusively by geometry.

If particles produced in the target are required to decay inside the detector, then the event rate scales with the fiducial volume. Taking a 40-ton LArTPC as representative, we find an approximately $28~{\rm m}^3$ volume for the DUNE ND as compared to $15~{\rm m}^3$ for the proposed bubble chamber.  As this loss in detector volume is compensated by the larger flux of BSM particles (due to the geometry discussed above), the number of BSM decays in the bubble chamber would be roughly the same as in the DUNE ND LArTPC. Advantageously, the neutrino scattering backgrounds would be under relatively tight theoretical control with no missing momentum due to nuclear recoil. Some authors have explored using additional elements of the DUNE ND complex to assist in reduced neutrino scattering backgrounds such as using the GArTPC \cite{Berryman:2019dme} or the DUNE-PRISM concept \cite{DeRomeri:2019kic}.

If instead BSM particles are detected via scattering signals (for instance millicharged particles are detected via their scattering on electrons \cite{Magill:2018tbb,Harnik:2019zee,ArgoNeuT:2019ckq}), then the event rate will scale like the detector's mass (instead of the detector's volume),  providing a substantial benefit to the LArTPC detector, and making the GArTPC strategy pursued in \cite{Berryman:2019dme} untenable. The hydrogen bubble chamber or a carbon subtracted measurement with SAND could provide a useful alternative in which statistics are reduced by roughly an order of magnitude, but wherein background systematics can be much better understood. 

The mass of BSM particles produced in the target can range anywhere from the sub-MeV scale to roughly $\sim 2$ GeV, a cutoff imposed by the masses of the $D$ mesons. Higher mass states can in principle be produced via, for instance, $\Upsilon$-mediated or Drell-Yan production \cite{Magill:2018tbb}, though the production cross sections are greatly reduced, and DUNE quickly loses its competitive reach when compared with LHC probes that pay a much smaller penalty.  

The above discussion assumes the beam is running in the mode needed for neutrino oscillation measurements.  Special runs may be taken with the magnetic focusing horns turned off, which would reduce the beam-neutrino induced backgrounds.  Alternatively, the target could be removed for a dedicated BSM data sample.  The proton beam would then travel to the hadron absorber at the end of the decay pipe, and prompt meson decays would occur there. The intensity of the beam would have to be reduced in this running mode in order not to damage the hadron absorber.

\subsubsection{Charged meson and muon progenitors} 
\label{sec:bsm:chargedmesonandmuonprogenitors}

Instead of producing BSM particles in the target or the hadron absorber, one may also consider production via the decays of long-lived charged mesons, i.e., $\pi^\pm$ and $K^\pm$, and their decay products i.e., $\mu^\pm$. For example, $\pi^\pm \rightarrow N \ell^\pm$, where $N$ is a heavy neutral lepton, is the most efficient production mechanism for heavy neutral leptons satisfying $M_N \leq m_\pi -m_\ell$. The parent mesons are focused by the magnetic horns, increasing the fluxes of their decay products (i.e., BSM particles and neutrinos) in downstream detectors.   Moreover, the mesons and muons propagate and decay in the $194$~m decay pipe, placing their decays in much closer proximity to the proposed deuterium hall compared to the DUNE ND hall. Muons have not been studied extensively in the literature, however a sizeable flux will range out in the dirt behind the muon alcove and they could provide sizeable fluxes of light particles such as axions in their decays. Indeed, even forty years ago, muon bremsstrahlung was identified as an efficient source of axion production \cite{Tsai:1986tx,}. The closer proximity to the decay pipe's terminus and the muon alcove can increase the flux of BSM particles at the deuterium bubble chamber by a factor of roughly five as compared to the DUNE ND complex, which easily compensates for the reduced fiducial volume of the bubble chamber as compared to the ND LArTPC. This would be an advantageous tool with which to search for muon-philic exotic particles such as an $L_\mu-L_\tau$ coupled $Z'$ or muon-coupled axion-like particle. 

An interesting case study where the advantages of nucleon target material vs. nuclear target material is manifest is the neutral-current detection signal of light dark matter $\chi A/N \rightarrow \chi A/N + \pi^0$. In this scenario, dark matter $\chi$ is produced in meson decays, travels towards the detector, and interacts via a vector mediator with the hadronic target.  The authors of \cite{deNiverville:2016rqh} anticipated that this channel could be the most sensitive probe of light dark matter in MiniBooNE whereas the actual experimental analysis employed on a CC$0\pi$ and a NC$0\pi$ analysis which relied on models of intranuclear pion absorption models \cite{MiniBooNE:2017nqe}. In a bubble-chamber experiment, the analysis of \cite{deNiverville:2016rqh} can be applied directly, and provides a concrete example of a BSM signature that would benefit from the absence of a complicated nuclear environment.

\subsubsection{Neutrino-induced BSM particle production } 

These events will be produced by neutrinos in the rock/dirt surrounding the detector or within the detector itself. If the differential upscattering cross section is forward-peaked and the lifetime of the BSM particles is long, then upscattering in the column of dirt between the detector and the target can be an efficient production mechanism for BSM particles. A popular model in which precisely this kind of behavior is expected features a transition magnetic dipole operator that connects a left-handed SM neutrino to a right-handed partner \cite{Magill:2018jla,Plestid:2020vqf,Schwetz:2020xra,Ismail:2021dyp,Brdar:2020quo,Jodlowski:2020vhr}. Alternatively, if the lifetime of the unstable BSM particle is short, then its detection necessarily requires that it be produced close to or inside of a neutrino detector. These kinds of models feature prominently in the context of the MiniBooNE low-energy excess \cite{Bertuzzo:2018itn,Vergani:2021tgc}. 

Depending on the nature of the scattering cross section, these models may predict very different scattering rates, topologies and kinematics when comparing hydrogen vs. argon. If the initial scattering is hard, then it may be sufficient in a LArTPC to simply correlate the two events which will be separated by a visible gap~\cite{Coloma:2017ppo,Atkinson:2021rnp}. Alternatively, the upscattering may involve relatively little energy or momentum transfer to the target nucleon/nucleus. For instance the model in \cite{Bertuzzo:2018itn} relies on a coherent upscattering mechanism for which both the nuclear and nucleon-level upscattering cross sections are known with high precision. A LArTPC then has numerous benefits since the cross section is coherently enhanced $\sigma \sim Z^2 \sigma_0$ and LArTPCs are highly capable of resolving $e^+e^-$ pairs. Because of the correlated events, background mitigation is often possible in the BSM scenarios sketched above and it is likely the case that for neutrino-induced BSM particle production inside the detector, the DUNE ND complex will outperform a dedicated search with a bubble chamber owing to its superior statistics.

\subsubsection{Carbon-subtracted data vs bubble chamber data } 

A relatively high-purity sample of CCQE events on hydrogen can be obtained from a hydrocarbon target provided the contribution from the carbon nuclei can be reliably subtracted.  The hydrocarbon target material proposed for the SAND detector contains two hydrogen atoms for each carbon atom.  As described in Section~\ref{sec:SAND}, the SAND detector will include both hydrocarbon radiators and pure carbon radiators whose neutrino scattering events can be subtracted from the hydrocarbon events using vertex reconstruction alone~\cite{Duyang:2018lpe,Petti:2019asx}. To further purify the sample of scatters on hydrogen before the subtraction, a
proposal using transverse kinematic imbalance (TKI) has been put forth which may be able to isolate a hydrogen-rich subsample~\cite{Lu:2015hea,Hamacher-Baumann:2020ogq}.

The caveat to this TKI-proposal is that it assumes that the scattering kinematics involve SM initial and final states, and that any deviations from the associated kinematic constraints are due to nuclear effects. All TKI variables are essentially functions of $\delta \vb{p}_T$, the residual transverse momentum of the lepton + nucleon system.  If an undetected BSM particle is produced in a neutrino scattering event on hydrogen and this BSM particle has a nonzero momentum transverse to the beam, then it will generate substantial TKI and the event will be labeled as a ``carbon-like'' scattering event. The TKI algorithm will thus systematically discard certain BSM signatures in which light BSM particles are produced on shell (see e.g.\ \cite{Berryman:2018ogk,Kelly:2019wow,}). 

A bubble chamber, by way of contrast, would be very sensitive to deviations from free-particle kinematics, independent of any cross section uncertainties. It would be limited entirely by detector resolution issues. In this respect, a bubble chamber design may be superior if the BSM signal can be discriminated from SM background using kinematics alone. For example, if the neutrino has initial state radiation of some light scalar particle~\cite{Berryman:2018ogk}, then using TKI variables in a bubble chamber one could reduce SM backgrounds to extremely low levels while maintaining reasonable signal purity. If, instead, one is interested in a signal which will pass free-particle TKI cuts with high efficiency, then the carbon-subtracted data set from the hydrocarbon components in SAND may be sufficient. 

\subsubsection{Exact kinematics and elementary targets \label{exact-kins} } 

In addition to well-understood hadronic inputs for neutrino cross section predictions, the use of an elementary nucleon target also offers basic kinematic advantages. Any neutrino experiment must contend with flux uncertainties due to the broad-spectrum nature of Fermilab's neutrino beams. When scattering on a nuclear target, this is a formidable challenge because the neutrino energy cannot be uniquely reconstructed; a recoiling nucleus can always carry away missing momentum and neutrons can hide missing energy. An elementary target does not suffer from these deficiencies. 

For example the anti-neutrino energy in the reaction $\bar{\nu} p \rightarrow  \ell^+ n$ can be uniquely reconstructed using the outgoing lepton kinematics. Likewise, the anti-neutrino energy in $\bar{\nu} p \rightarrow \bar{\nu} p$ can be uniquely reconstructed using  the final state proton's kinematics. The observable $R_{\rm CC:NC}=N_{\rm CC}/ N_{\rm NC}$ can then be constructed for different neutrino energies and is directly related to the charged-current to neutral-current ratio  $R_{\rm CC:NC}= \sigma_{\rm CC} / \sigma_{\rm NC}$ as a function of neutrino energy. 

\subsubsection{Outlook} 

A hydrogen/deuterium bubble chamber offers the realistic possibility of probing neutrino nucleon interactions with unprecedented percent-level precision. More work must be done to understand the relative strengths and weaknesses of different detectors relative to the DUNE ND complex and in particular the tradeoffs between bubble-chamber data and carbon-subtracted data from SAND. Generic expectations are that a bubble chamber can provide similar event yields to the ND LArTPC, but with much lower theoretical uncertainties (many of which are not accounted for in BSM studies in the literature). Further research into detector performance will be required to quantify the signal-to-background ratios in the two experiments, however it is important to emphasize the role of theoretical uncertainties in any such comparison. Provided the hydrogen/deuterium bubble chamber offers competitive (even if slightly worse) detector performance, then the absence of large theory uncertainties will likely result in stronger constraints for any model which is limited by systematic uncertainties stemming from neutrino-nucleus scattering. 

While we have focused most of our discussion on a comparison between the DUNE's ND-LAr + ND-GAr/TMS, a bubble-chamber detector, and a carbon-subtracted dataset obtained from the SAND, it is worth emphasizing that hydrogen data will also complement a LArTPC-based BSM program. Broadly speaking, the DUNE ND and the hydrogen data can be used as sidebands for one another. In particular, they can serve to test hypotheses about potential BSM signal origins. For instance, if an excess of $e^+e^-$ pairs is seen in the ND complex, and the purported explanation is dark matter produced from prompt meson decays in the target (i.e.\ beam stop), then the flux of dark matter passing through the bubble chamber would be uniquely fixed, and so will the rate of $e^+e^-$ deposition. In this way the bubble chamber can be used as a direct test of new physics interpretations within the DUNE ND complex.

\subsection{Impact on precision measurements and hadronic physics
}  \label{sec2_6:impact_on_hadron_physics}

The $\nu$D and $\nu$H scattering processes
play a dual role of determining elementary amplitudes,  and validating
nuclear and radiative corrections in the simplest nuclei. 
We focus here on two precision observables that demonstrate the impact of new elementary target neutrino data beyond the neutrino oscillation program: first, the structure function $F_3$ for CC inclusive charged current neutrino scattering, to determine radiative corrections for $V_{ud}$ extractions from beta decay; second, the nucleon charged current axial radius $r_A^2$.

\subsubsection{Nuclear beta decay and CKM unitarity}  \label{sec2_6_1:CKM_and_beta_decay}

\begin{figure}[ht]
    \centering
    \includegraphics[width=0.5\textwidth]{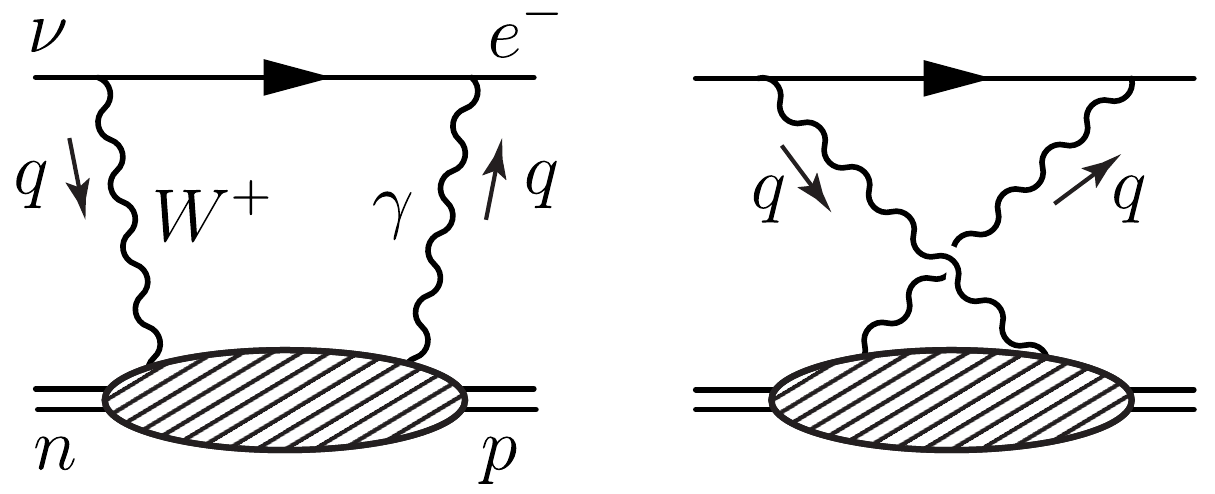}
    \caption{(From Ref.~\cite{Seng:2018yzq}) The $\gamma W$-box diagram contributing to neutron $\beta$-decay.}
    \label{fig:box}
\end{figure}

In the SM, the Cabibbo-Kobayashi-Maskawa (CKM) quark mixing matrix is unitary, with, e.g., the top-row constraint $\Delta_u=|V_{ud}|^2+|V_{us}|^2+|V_{ub}|^2-1=0$. However, a combination of world data \cite{ParticleDataGroup:2020ssz} suggests a $2\sigma$- or $3\sigma$-deficit, $\Delta_u=-0.0015(7)$ or $\Delta_u=-0.0015(5)$, depending on nuclear uncertainties employed in the extraction of the top-left corner element $V_{ud}$ \cite{Gorchtein:2018fxl,Hardy:2020qwl}.
The CKM matrix element $V_{ud}$ is extracted from a $\beta$ decay process $I$ according to the master formula,
\begin{align}
    |V_{ud}^I|^2=\frac{K^I}{1+\Delta_R^V},\quad \mathrm{with}\quad
    \begin{array}{l}
         K^{0^+-0^+}=0.97154(22)_{\rm exp,\,nucl}(54)_{\rm NS} \\
         K^\mathrm{free\,n}=0.97149(66)_{\tau_n}(169)_{g_A} 
    \end{array}
\end{align}
with the numerator $K^I$ summarizing process-specific measurements and corrections. The first (second) uncertainty in the factor $K$ above refer to a combination of experimental and nuclear correction $\delta_C$ (nuclear structure correction $\delta_\mathrm{NS}$) for superallowed $0^+-0^+$ nuclear decays, and to neutron lifetime $\tau_n$ (axial charge $g_A$) for the free-neutron decay, respectively.
The universal electroweak radiative correction (EWRC) $\Delta_R^V$ is common to both decays and contains the $\gamma W$-box diagram depicted in Figure~\ref{fig:box}, 
\begin{align}
    \Delta_R^V=2\Box_{\gamma W}+\mathrm{model\;independent},
\end{align}
which has long been identified as the main source of the theoretical uncertainty \cite{Sirlin:1977sv,Marciano:1985pd}.
Recently, this correction has been reevaluated within the dispersion relation framework \cite{Seng:2018yzq,Seng:2018qru} with the result $\Delta_R^V= 0.02467(22)$, significantly larger than the previous evaluation $\Delta_R^V= 0.02361(38)$ \cite{Marciano:2005ec} and about twice as precise. The new result was independently confirmed by three other groups~\cite{Czarnecki:2019mwq,Shiells:2020fqp,Hayen:2020cxh}.

In the dispersion framework, the $\gamma W$-box correction evaluated at zero energy and momentum transfer is expressed as 
\begin{equation}
\Box_{\gamma W}=\frac{3\alpha}{2\pi}\int_0^\infty \frac{dQ^2}{Q^2} \frac{M_W^2}{M_W^2+Q^2}M_3^{(0)}(1,Q^2),\label{eq:boxNachtmann}
\end{equation}
where $M_3^{(0)}(1,Q^2)$ is the first Nachtmann moment of the structure function $F_3^{(0)}$~\cite{Nachtmann:1973mr,Nachtmann:1974aj}
\begin{equation}
M_3^{(0)}(1,Q^2)=\frac{4}{3}\int_0^1 dx \frac{1+2r}{(1+r)^2}F_3^{(0)}(x,Q^2),
\label{eq:NachtmannDef}
\end{equation}
and $r=\sqrt{1+4M^2x^2/Q^2}$. The inclusive nucleon structure function $F_3^{(0)}$ describes the hadronic blob in Figure~\ref{fig:box} and arises from the time-ordered product of the isovector charged axial current and the isoscalar electromagnetic current. As such, this structure function is not an observable. By isospin symmetry and minimal modeling, however, it can be related to its purely charged-current counterpart measurable via a combination of neutrino and antineutrino data:
\begin{align}
    \frac{d^2\sigma^{\nu p}}{dx dy}-\frac{d^2\sigma^{\bar\nu p}}{dx dy}\propto F_3^{\nu p+\bar\nu p}=(F_3^{W^+}+F_3^{W^-})/2.
\end{align}
The first Nachtmann moment $M_3^{\nu p+\bar\nu p}(1,Q^2)$ of $F_3^{\nu p+\bar\nu p}$ has been extensively studied with the old bubble-chamber experiments in the context of the Gross-Llewellyn-Smith sum rule which requires $M_3^{\nu p+\bar\nu p}(1,Q^2)\to3$ for $Q^2\to\infty$. 
\begin{figure}[ht]
    \centering
    \includegraphics[width=0.5\textwidth]{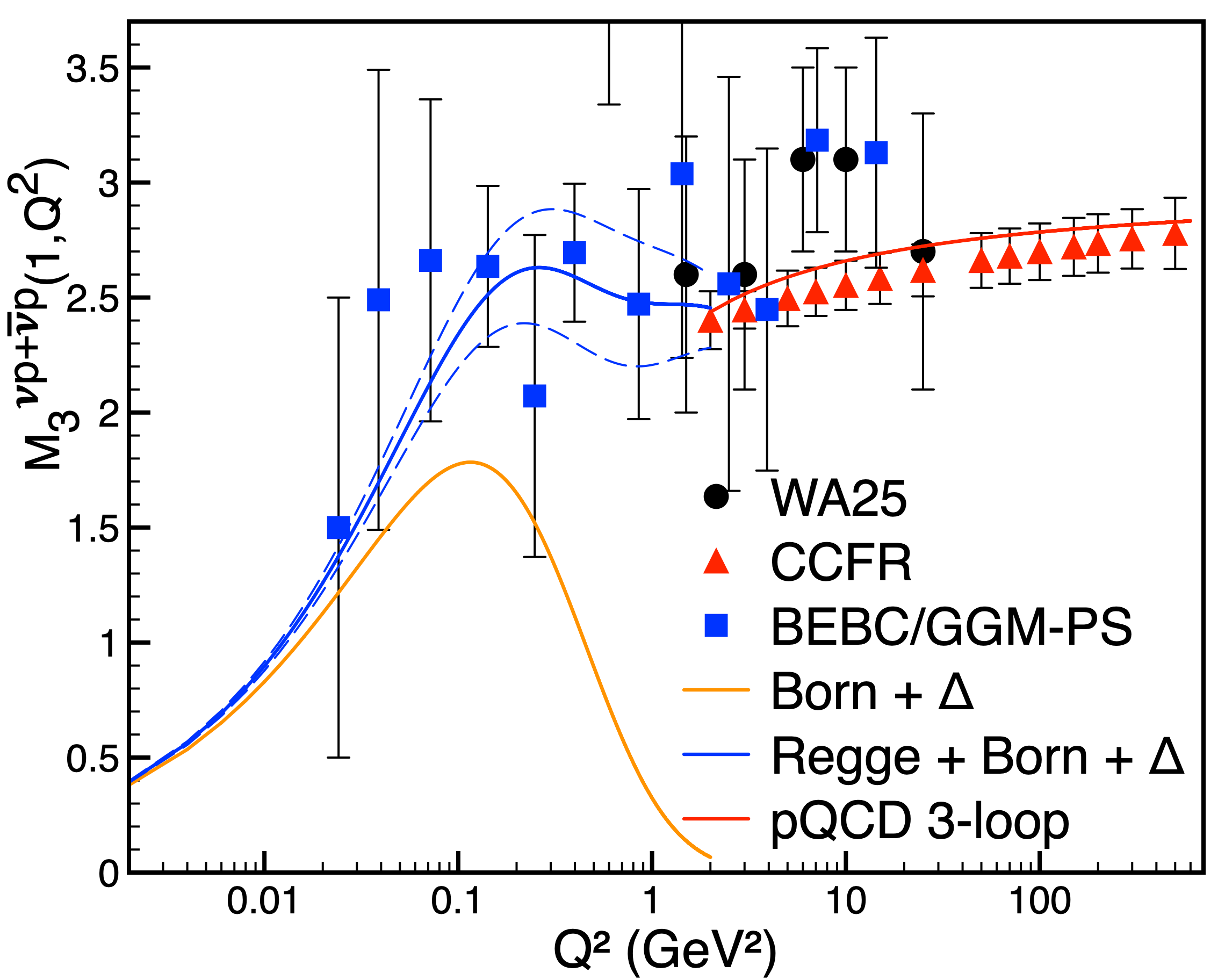}
    \caption{World data on the first Nachtmann moment $M_3^{\bar\nu p+\nu p}(1,Q^2)$ of the CC inclusive structure function $F_3$ as a function of $Q^2$ along with theoretical curves (see text for explanation).}
    \label{fig:M3}
\end{figure}
World data on this sum rule as a function of $Q^2$ from WA25~\cite{Bolognese:1982zd}, BEBC/Gargamelle~\cite{Allasia:1985hw} and CCFR~\cite{Kim:1998kia,Kataev:1994rj} collaborations are displayed in Figure~\ref{fig:M3}. Also displayed is the perturbative QCD (pQCD) prediction at 3-loop~\cite{Larin:1991tj} (red curve) and the phenomenological low-$Q^2$ model of Refs.~\cite{Seng:2018qru,Seng:2018yzq} (blue curve) operating with the dominant elastic, $\Delta(1232)$ and Regge contributions. Elastic (Born) contribution described by the nucleon form factors dominates $M_3^{\nu p+\bar\nu p}(1,Q^2)$ at $Q^2\lesssim0.01$~GeV$^2$, that of $\Delta(1232)$ for 0.01~GeV$^2\lesssim Q^2\lesssim0.1$~GeV$^2$, 
the sum of the two is shown by orange curve in Figure~\ref{fig:M3}. 
The Regge contribution starts dominating above that low $Q^2$ region and smoothly joins the pQCD curve at $Q^2=2$~GeV$^2$. The dashed blue curves display the uncertainty obtained by fitting the Regge contribution to the data in the intermediate region. Note that the low-$Q^2$ inclusive data from BEBC/Gargamelle were not used to fit Born and $\Delta(1232)$ contributions; rather, more modern neutrino data that constrain the axial nucleon form factor and the $N\to\Delta$ transition were used. Because these data only cover limited high-$x$ portion of the integral in Eq.~(\ref{eq:NachtmannDef}) they are not shown on the plot.

In this approach, the experimental precision of inclusive $\nu/\bar\nu$ data directly translates into the uncertainty of $\Delta_R^V$. Specifically, a better determination of the first Nachtmann moment $M_3^{\nu p+\bar\nu p}(1,Q^2)$ for $Q^2\leq2$ GeV$^2$ and of the strength of the Regge contribution will help reducing the uncertainty of $\Delta_R^V$. The path from 
$M_3^{\nu p+\bar\nu p}(1,Q^2)$ to $M_3^{(0)}(1,Q^2)$ which gives $\Delta_R^V$ is straightforward but retains some model dependence: it requires a decomposition of the inclusive low-$Q^2$ data into elastic, resonance, Regge contributions with definite isospin quantum numbers, and each contribution is then isospin rotated to obtain its isoscalar counterpart. To test the model dependence of this procedure, a lattice QCD calculation of the first Nachtmann moment $M_3^{(0)}(1,Q^2)$ was performed on the pion~\cite{Feng:2020zdc}, which was then translated into that on the nucleon~\cite{Seng:2020wjq}. 
\begin{figure}[ht]
    \centering
    \includegraphics[width=0.5\textwidth]{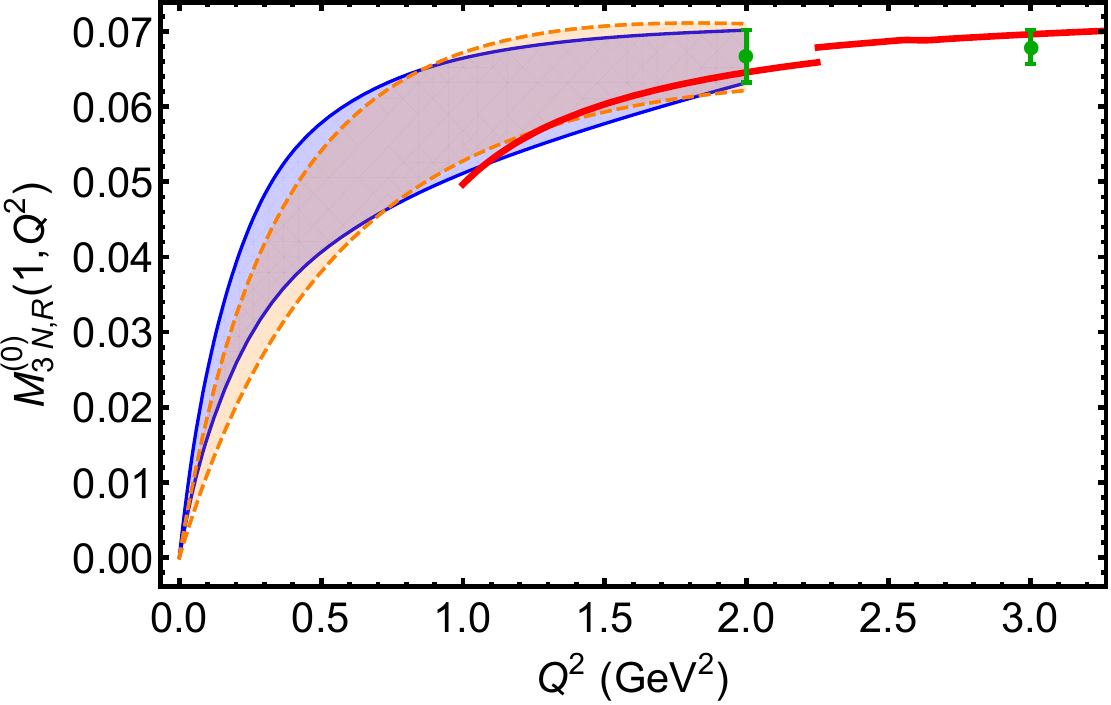}
    \caption{Regge contribution to $M_3^{(0)}(1,Q^2)$ on the nucleon: solid blue curves depict upper and lower bounds of the lattice QCD-based result, dashed orange curves those of the phenomenological data-driven result. The shaded regions between the curves represent the respective uncertainty. Solid green data points are the isoscalar analog of the lowest $Q^2$ CCFR data on $M_3^{\nu p+\bar\nu p}(1,Q^2)$, and the solid red curve is the N$^3$LO pQCD prediction.}
    \label{fig:M3LQCDPheno}
\end{figure}
A comparison of the phenomenological data-based and lattice-QCD based estimates of $M_3^{(0)}(1,Q^2)$ is displayed in Figure~\ref{fig:M3LQCDPheno}, demonstrating a very close agreement between the two, both in the central value and the uncertainty. Putting further high-precision data on this plot will allow for a quantitative comparison with the lattice QCD.

Another note is in order here: the data in Fig.~\ref{fig:M3} were not obtained on a proton target but are derived from those on light and intermediate nuclei upon applying nuclear corrections. CCFR data were taken on a Fe target, and existing deuteron data suffer from large uncertainties. So, new, high-precision hydrogen and deuterium data, free from complicated nuclear corrections, will largely improve the situation. 

Moreover, nuclear modifications of single-nucleon structure functions are of interest in their own right: the nuclear structure correction to the nuclear $\beta$ decay rate, $\delta_{\rm NS}$, is obtained by subtracting the free-neutron $\gamma W$-box from that evaluated on a nucleus~\cite{Seng:2018yzq}.  Independent data on H, D and $^{12}$C provides a way to disentangle $\delta_{\rm NS}$  and $\Delta_R^V$ with a controlled, data-driven uncertainty. 

\subsubsection{Nucleon axial radius} \label{sec2_6_2:axial_radius}

\begin{figure}[ht]
    \centering
    \includegraphics[width=0.5\textwidth]{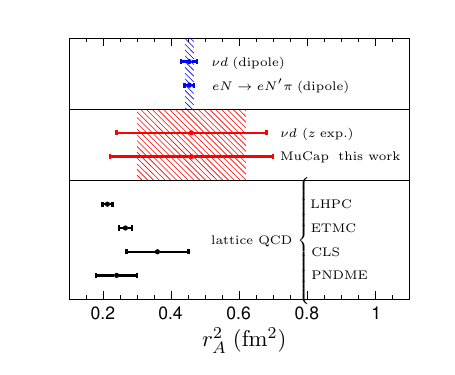}
    \caption{(From Ref.~\cite{Hill:2017wgb})  Squared axial radius determined by different processes.
    Top: Dipole fits to $\nu d$ and $eN \to eN^\prime \pi$ data employed before 2016 resulted
    in an average \raq\ with small uncertainty (hatched blue
    band)~\cite{Bodek:2007ym}.
    Middle: Replacing the unjustified dipole assumption with the $z$ expansion allowed a model-independent
    extraction of \raq\ from $\nu d$ data and increased the uncertainty. The hashed red region represents
    the best average from this work combining the new determination from MuCap with the $\nu d$ result.
    Bottom: Early lattice QCD results denoted
    by their collaboration acronyms. Several calculations
    tend towards lower values of \raq\ compared
    to the historical dipole average.}
    \label{fig:rA2}
\end{figure}

The nucleon axial radius $r_A^2$ [cf. Eq.~(\ref{eq:rA2})] is a fundamental parameter describing the nucleon.   
The analogous vector radius of the proton can be probed using a variety of processes including regular and muonic hydrogen spectroscopy, electron-proton scattering, and first principles lattice QCD.  
Similarly, $r_A^2$ can be probed using several processes, as illustrated in Fig.~\ref{fig:rA2}.  

As discussed in Sec.~\ref{sec2_2_1:lattice_QCD}, new elementary target data is needed to resolve tensions between different extractions. 
In a simplified description, the statistical error of the axial radius scales with the number $N$ of CCQE interaction events as $1/\sqrt{N}$:
\begin{align}\label{eq:rA2error}
  \delta r_A &=  \left(\delta  r_A \right)_0  \sqrt{\frac{N_0}{N}}\,, \qquad \left(\delta  r_A \right)_0 \sim 0.2-0.25~\mathrm{fm} , \qquad N_0 = 10^4 \,,
\end{align}
where $(\delta r_A)_0$ and $N_0$ are representative of the existing ANL and BNL datasets. To obtain the error estimate (\ref{eq:rA2error}), default fits for the vector and axial form factors from Refs.~\cite{Borah:2020gte} and~\cite{Meyer:2016oeg} were used to produce pseudo-random set of events. The same vector form factors from~\cite{Borah:2020gte} were then used to fit the simulated data using a 4-parameter z-expansion fit form for the axial form factor.

\subsection{The importance of deuterium data}
\label{sec:needfordeuterium}

A deuterium target provides a source of neutrons in a relatively simple and loosely-bound nucleus.  
It is important to recall that neutrino charged-current quasielastic scattering, the primary interaction of many oscillation experiments, can only proceed with a neutron in the initial state:
a direct study of this important process needs a neutron target. Furthermore, as this important neutrino neutron interaction is a fundamental input to models of neutrino-nucleus interactions in neutrino event generators, it is of essential importance to measure separately, including radiative corrections and inelastic channels that are not simply related, e.g. via isospin relations, to the neutrino proton interaction.   

\subsubsection{Neutron targets and extraction of parton distribution functions}
At the higher end of the $W$ spectrum, the need for a deeper study of QCD and associated extraction of nucleon parton distribution functions (PDFs) is another strong argument for having a large sample of neutrino-neutron statistics.
 
For brevity, the following is based on a set of simplifying assumptions. These include using leading-order expressions for the cross sections, using asymptotic expressions for the kinematics, ignoring the longitudinal cross section, and in some cases using $s=\bar s$ and $c=\bar c$.
 With these simplifications one can write the cross section expressions directly in terms of the various PDFs. These are as follows:
    
  \begin{align}
      \frac{d\sigma^{\nu p}}{dx dy} \equiv \sigma^{\nu p} &\propto [d + s] + [\bar u + \bar c] (1-y)^2 \,,
\nonumber \\
  \sigma^{\bar \nu p} &\propto [u + c](1-y)^2 + [\bar d + \bar s] \,,
\nonumber \\
      \sigma^{\nu n} &\propto [u + s] + [\bar d +\bar c](1-y)^2 \,,
\nonumber \\
      \sigma^{\bar \nu n} &\propto [d + c](1-y)^2 + [\bar u + \bar s] \,.
  \end{align}
Some additional notation that will be useful is the following:
    \begin{equation}
        \Sigma^p \equiv \sigma^{\nu p} + \sigma^{\bar \nu p} 
        \,, \qquad 
        \Delta^p \equiv \sigma^{\nu p} - \sigma^{\bar \nu p}
    \end{equation}
and similarly for a neutron target.
    
The first thing to note is that at large values of $x$ where one can neglect all but the $d$ and $u$ quarks, $\sigma^{\nu p}$ constrains the $d$ quark PDF while $\sigma^{\bar \nu p}$ constrains the $u$ PDF. Therefore, these proton cross sections will constrain the $d/u$ ratio. The same information is available with a neutron target, but nothing new is added. Thus, to justify the use of a neutron target for the extraction of nucleon PDFs one must examine the region of low to moderate values of $x$ where all flavors of PDFs must be considered.
    
Now, by assuming $s=\bar s$ and $c=\bar c$ one can see that 
    \begin{equation}
    \Delta^p \propto [d - \bar d] - [u - \bar u](1-y)^2 
    \,, \qquad
    \Delta^n \propto [u - \bar u] - [d - \bar d] (1-y)^2.
\end{equation}
Rearranging, one has
\begin{equation}
    \Delta^p + (1-y)^2 \Delta^n \propto [d - \bar d] [1-(1-y)^4] 
    \,, \qquad
    \Delta^p(1-y)^2 + \Delta^n  \propto [u - \bar u][1-(1-y)^4]
\end{equation}
Therefore, the use of neutron and proton data together allows one to extract both the $d$ valence and $u$ valence PDFs. 

Next, consider
\begin{equation}
    \Sigma^p-\Sigma^n \propto [d + \bar d - u - \bar u][1-(1-y)^2]
\end{equation}
which shows that given the valence $u$ and $d$ PDFs one can, with the neutron input, extract the quantity $\bar d -\bar u$. This can also be extracted from $pp$ and $\bar p p$ lepton pair production data, so the neutrino neutron data provides a complementary determination of the difference between the $ \bar d$ and $\bar u$ PDFs.  

Finally, consider
\begin{equation}
    \Sigma^p + \Sigma^n \propto [u + \bar u + d + \bar d + 2s + 2 \bar s] + [u + \bar u + d + \bar d + 2c + 2 \bar c](1-y)^2.
\end{equation}
Given a perturbative estimate of $c + \bar c$  one could then extract $s + \bar s$. This information would complement the information on the strange PDF coming from neutrino di-muon production as well as collider data for $W$-boson production.

The above equations, based on a set of simplifying assumptions, serve to illustrate the power of having both neutron and proton target neutrino data. A realistic global PDF analysis based on NLO or NNLO expressions with correct kinematics would be able to constrain the PDF combinations noted above, thereby providing both new information and cross checks for the continued study of flavor differentiation of nucleon PDFs.

\subsubsection{Neutron targets and hadronization studies}

The application of hadronization gives the multiplicity and kinematics of hadrons before final state interactions and consequently impacts the estimation of backgrounds and calorimetric energy reconstruction.  

A recent publication regarding hadronization tuning in GENIE V.3~\cite{GENIE:2021wox} emphasizes the very different hadronization tuning for $\nu$-neutron scattering as opposed to $\nu$-proton scattering and the fact that both proton and neutron hadronization studies had to be based on the low statistics and associated systematics of bubble-chamber studies from the 1980s. Improving the hadronization studies for proton targets only and neglecting the hadronization study of the neutron targets, which dominate neutrino-nucleus interactions, will leave the simulated hadronization of $\nu$ - A in a quite imprecise and uncertain state.

\section{Experimental options \label{sec:expt}}
\subsection{The DUNE near detector}

\begin{figure}[ht]
    \centering
    \includegraphics[width=0.95\textwidth]{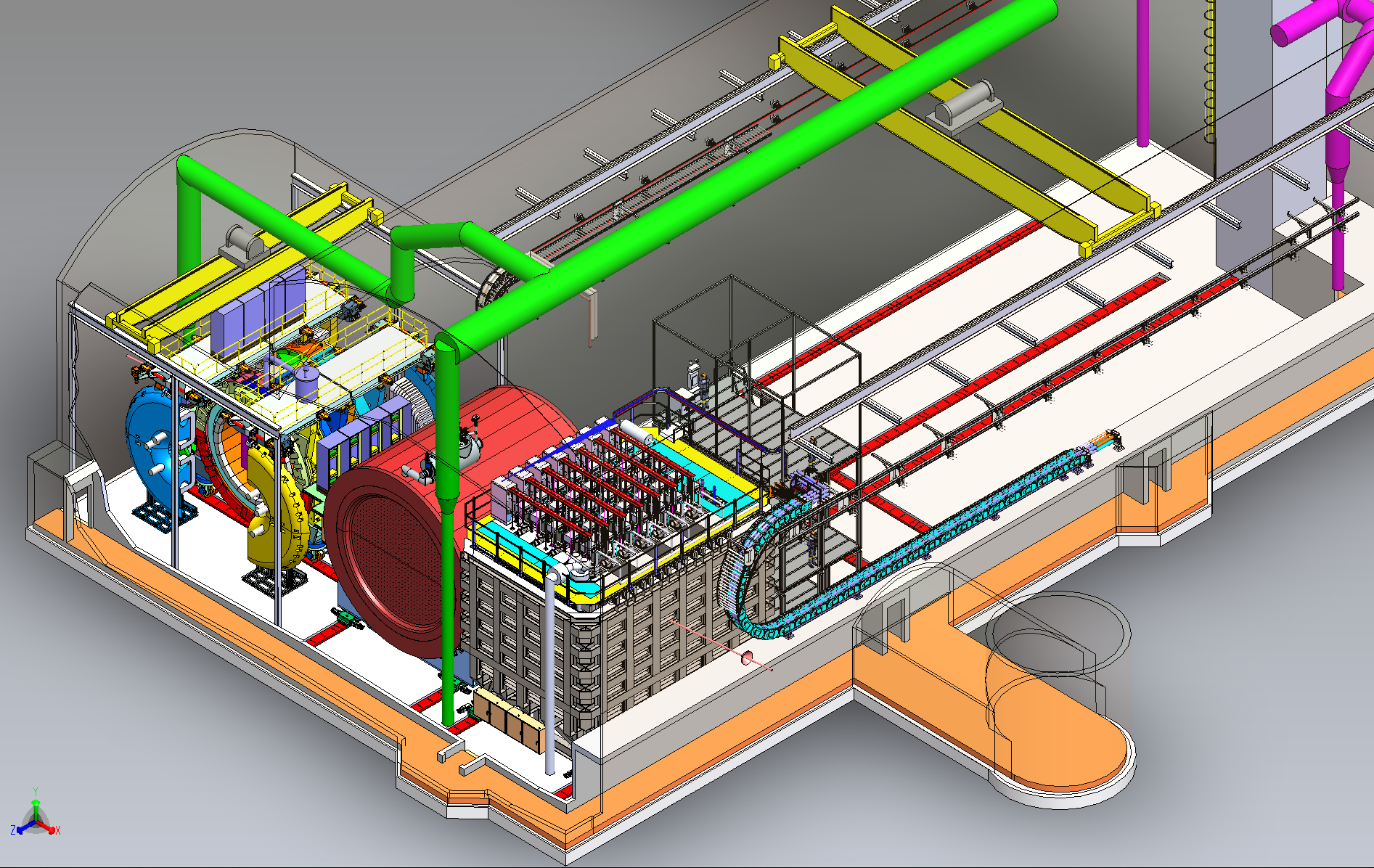}
    \caption{Schematic of the DUNE near detector complex.}
    \label{fig:NDhall}
\end{figure}

The DUNE near detector complex~\cite{DUNE:2021tad} will be located on the Fermilab site approximately 574~m from the target.  The reference design consists of three components: ND-LAr, which is a liquid argon time projection chamber (TPC) with pixelated readout (the ArgonCube technology), a multi-purpose detector consisting of a high-pressure (10~bar) gaseous argon TPC surrounded by an electromagnetic calorimeter in a 0.5~T magnetic field (called ND-GAr), and finally a beam monitor that will re-purpose the KLOE magnet and calorimeter (called SAND).  ND-LAr and ND-GAr will be able to move off axis in order to take data with different energy neutrino beams in order to implement the PRISM (Precision Reaction-Independent Spectrum Measurement) methodology.  The beam monitor is called SAND (System for on-axis Neutrino Detection) and it remains on-axis at all times.  A schematic of the near detector hall and the three detector components is shown in Figure~\ref{fig:NDhall}.

The reference design described above is, at the time of writing, not envisaged as the suite of near detector components that will be installed when the LBNF beam turns on.  Instead, a staged approach is proposed, in which ND-GAr is reserved for a future upgrade, and the measurement of muons escaping from ND-LAr is performed by a magnetized steel spectrometer called TMS, which moves with ND-LAr as part of the PRISM measurement.  Muon energies are measured in TMS using their ranges in the steel stack, which has scintillating detector layers between the steel layers to measure particle trajectories.  The steel is magnetized in order to determine the signs of the charges on the muons.

\subsubsection{A hydrogen-target detector component}

The contents of the near detector hall at LBNF will evolve over time as resources and materials become available.  Given the interest in measuring neutrino interactions on H$_2$ and D$_2$, one might consider adding a large pure-hydrogen gas TPC or liquid hydrogen/deuterium bubble chamber to the near detector hall.  It would be a very ambitious project to do this, however.  Although the facility could support such a detector from a technical standpoint, the cost of doing so would be prohibitive due to fire-protection underground safety codes.  As per the National Fire Protection Association's codes, any underground facility which stores more than 40~kg of a flammable liquid or gas would have to be built under explosion-proof electrical guidelines.   In order to get sufficient statistics to meet the scientific goals, a much larger target is required. This would require the entire near detector hall to be built under the explosion-proof electrical guidelines.  The cost would be excessive and the impact on the core mission of the facility would be adversely effected.  Therefore, we conclude that this is not a viable path forward.

\subsubsection{Measuring neutrino-hydrogen interactions in SAND}
\label{sec:SAND} 

The SAND detector is expected to be part of the DUNE ND complex from the beginning of the data taking (``Phase 1"). 
The inner tracker of SAND is based on a Straw Tube Tracker (STT) which was designed to provide a control of the configuration, chemical composition and mass of the neutrino targets comparable to electron scattering experiments~\cite{Petti:2019asx}. 
The targets are provided by multiple thin -- 1-2 \% of radiation length -- planes of passive materials, interleaved by active straw layers of negligible mass (about 3\% of the total mass). The targets can be removed or replaced with different materials during data taking and account for about 97\% of the total detector mass. The average density can be tuned in the range $0.005 \leq \rho \leq 0.18$ g/cm$^3$ and the total detector thickness is comparable to one radiation length. These latter features are critical in achieving an accurate reconstruction of the event kinematics in a plane transverse to the beam direction from a precise measurement of the four-momenta of the final state particles.
We note that a similar low-density concept was previously used by the NOMAD experiment, which was explicitly designed to exploit the transverse plane kinematics for event selection~\cite{Astier:2001yj}. The momentum scale is expected to be calibrated to about 0.2\% using reconstructed $K_0 \to \pi^+\pi^-$ and $\Lambda \to p \pi^-$ decays~\cite{Wu:2007ab,Duyang:2019prb}.

\begin{table}[tb] 
\centering
\caption{Number of events per year expected to be selected in SAND with the default low-energy FHC and RHC beams for the various topologies in $\nu(\bar \nu)$-H CC interactions.}
\vspace{0.3 cm}
\label{tab:SAND-evts} 
\begin{tabular}{|l|r|} \hline
             &  H selected \\
CC process    &  Events/year \\ \hline\hline
$\nu_\mu p \to \mu^- p \pi^+$   &   408,000 \\
$\nu_\mu p \to \mu^- p \pi^+ X$   &  152,000 \\
$\nu_\mu p \to \mu^- n \pi^+ \pi^+ X$   &  19,000  \\
\hline
$\nu_\mu$ CC inclusive on H~~  & 579,000 \\
\hline\hline
$\bar \nu_\mu p \to \mu^+ n$   &    172,000 \\
$\bar \nu_\mu p \to \mu^+ p \pi^-$   &   61,000 \\
$\bar \nu_\mu p \to \mu^+ n \pi^0$   &   42,000 \\
$\bar \nu_\mu p \to \mu^+ p \pi^- X$   &  27,000 \\
$\bar \nu_\mu p \to \mu^+ n \pi \pi X$   & 31,000 \\
\hline
$\bar \nu_\mu$ CC inclusive on H~~ & 333,000   \\
\hline
\end{tabular}
\end{table} 

The control of the targets offered by the STT in SAND enables the implementation of a ``solid" hydrogen concept in which $\nu(\bar \nu)$ interactions on H are obtained from the subtraction of pure polypropylene (CH$_2$) and graphite (C) targets~\cite{Petti:2019asx}. The technique is conceived to be model-independent, as the two target materials are interleaved throughout the STT volume to guarantee the same acceptance.
The total amount of hydrogen embedded within the STT fiducial volume in SAND is roughly equivalent to about 10~m$^3$ of liquid H$_2$. The purity of the selected CC H samples can be enhanced with a kinematic analysis taking advantage of the excellent reconstruction of the transverse plane variables in STT~\cite{Duyang:2018lpe,Duyang:2019prb}. This kinematic selection can be applied to all available inclusive and exclusive topologies in both $\nu$ and $\bar \nu$ CC interactions on H and can result in purities up to 80-90\%, depending on the specific topology~\cite{Duyang:2018lpe}. 
Table~\ref{tab:SAND-evts} summarizes the corresponding numbers of events per year expected to be selected in SAND. Similar kinematic selections have been demonstrated by NOMAD~\cite{Astier:2001yj} in more critical background conditions. We note that the ``solid" hydrogen concept can still be applied with no or limited kinematic selection, at the price of a larger statistical uncertainty from the subtraction procedure. 

The high-statistics samples of $\nu$-H and $\bar \nu$-H CC interactions which are expected to be collected in SAND can provide a valuable tool to constrain the systematic uncertainties related to the (anti)neutrino flux and to the nuclear smearing in data originated from the various nuclear targets $A$ available in SAND (e.g. C, Ar, etc.)~\cite{Petti:2019asx}. Exclusive processes like the single pion $\nu_\mu p \to \mu^- p \pi^+$ and the quasi-elastic $\bar \nu_\mu p \to \mu^+ n$ on H at small energy transfer $\nu$ can be used to determine the relative $\nu_\mu$ and $\bar \nu_\mu$ flux as a function of energy with an accuracy of about 1\%~\cite{Duyang:2019prb}. A direct comparison of inclusive and exclusive event topologies from a nuclear target $A$ and from H within the same SAND detector can constrain the corresponding nuclear effects and help to calibrate the reconstructed neutrino energy scale.

\subsubsection{Hydrogen-rich gas options for ND-GAr}
ND-GAr, with its full solid-angle acceptance and MeV-level proton tracking threshold, is ideal for performing exclusive final-state measurements. In the exclusive channels, $\nu+\textrm{A}\rightarrow\ell+\textrm{hadrons}+\textrm{A}'$, where the final-state hadrons are measured, the details of the intranuclear dynamics of the interactions can be precisely extracted by exploring momentum conservation in the transverse plane to the neutrino direction, regardless of the unknown neutrino energy. This technique of measuring Transverse Kinematic Imbalance (TKI) in neutrino interactions~\cite{Lu:2015hea,Lu:2015tcr,Furmanski:2016wqo,Lu:2019nmf} has been successfully applied in MINERvA~\cite{MINERvA:2018hba,MINERvA:2019ope,MINERvA:2020anu} and T2K~\cite{T2K:2018rnz,T2K:2021naz}. With the state-of-the-art tracking resolution, ND-GAr will use TKI to further explore neutrino-argon interactions with unprecedented precision~\cite{DUNE:2021tad}.

One unique advantage of the ND-GAr, compared to other DUNE ND components, is its flexibility to
use various gas mixtures as interaction targets. Possible hydrogen-rich mixtures include binary systems such as He-Alkane. For example, in terms of the ratio between the numbers of neutrino-interacting free protons and bound protons, a mixture of 90\% He + 10\% CH$_{4}$ equals to polystyrene (CH), and  75\% He + 25\% CH$_4$ equals to polypropylene (CH$_2$); at 50\%, CH$_{4}$ provides the same target mass as pure hydrogen gas (H$_2$)---higher concentration will not only further improve the event purity, but also increase the free-proton target mass.

With the help of the superb tracking of ND-GAr, a successful search for a safe and hydrogen-rich gas mixture would enable measurements of
event-by-event neutrino-hydrogen interactions~\cite{Lu:2015hea,Hamacher-Baumann:2020ogq}. The idea was that beamed neutrinos  interacting on hydrogen could be selected out of those on the heavier targets via TKI once sufficient momentum resolution is achieved: with perfect tracking, interactions on hydrogen would have balanced final-state transverse momenta (that is, zero TKI), while the TKI on heavier nuclei is irreducibly wide due to nuclear effects such as Fermi motion and final-state interactions. For example, with the projected neutrino beam exposure at DUNE and the expected tracking resolution of ND-GAr that is as good as that of the ALICE TPC, selecting small-TKI events of the channels $\smash{\overset{\scalebox{.3}{(}\raisebox{-1.7pt}{--}\scalebox{.3}{)}}{\nu}}+\textrm{H}\rightarrow\ell^\mp+\textrm{p}+\pi^{\pm}$ could deliver neutrino-hydrogen events of the order of 10$^4$ per year with purity above 90\% using methane gas~\cite{Hamacher-Baumann:2020ogq}.

\subsection{A dedicated facility in the LBNF beamline }

\begin{table}[bt]
\begin{center}
\caption{Predicted events per metric ton per year for the LBNF beam ($1.1\times 10^{21}$ POT) on Hydrogen, FHC.  The processes are:  charged-current quasielastic (CCQE), charged-current resonant (CCRES), charged-current deep-inelastic scattering (CCDIS), and their neutral-current counterparts, NCQE, NCRES and NCDIS.}
\vspace{0.3 cm}
\label{tab:fhcH}
\begin{tabular}{|l|r|r|r|r|}
\hline
Process &  $\nu_e$ & $\nu_{\mu}$ & ${\bar\nu}_e$ & ${\bar\nu}_{\mu}$ \\ \hline
CCQE      &     0  &      0 & 1010 & 27500 \\
CCRES     &  5500  & 527000 & 890  & 18100 \\
CCDIS     &  8700  & 402000 & 2630 & 43500 \\
\hline
Total CC  & 14200  & 929000 & 4530 & 89100 \\ \hline \hline
NCQE  & 1120 & 111000     &  190 & 5300 \\
NCRES  & 2040 & 191000    &  360 & 8100 \\
NCDIS  & 4140 & 206000    &  810 & 13800 \\ \hline
Total NC  & 7300 & 508000 & 1360 & 27200 \\ \hline
\end{tabular}
\end{center}
\end{table}

\begin{table}[thb]
\begin{center}
\caption{Predicted events per metric ton per year for the LBNF beam ($1.1\times 10^{21}$ POT) on Deuterium, FHC.  The processes are:  charged-current quasielastic (CCQE), charged-current resonant (CCRES), charged-current deep-inelastic scattering (CCDIS), and their neutral-current counterparts, NCQE, NCRES and NCDIS.}
\vspace{0.3 cm}
\label{tab:fhcD}
\begin{tabular}{|l|r|r|r|r|}
\hline
process &  $\nu_e$ & $\nu_{\mu}$ & ${\bar\nu}_e$ & ${\bar\nu}_{\mu}$ \\ \hline
CCQE   &  2830    & 287000  &   394 &  9950 \\
CCRES  &  5590    & 521000  &   862 & 18200 \\
CCDIS  & 13000    & 592000  &  1990 & 32800 \\ \hline
Total CC  & 21400 & 1400000 &  3250  & 61000 \\ \hline\hline
NCQE  & 1100      & 110000  &   183  & 5170 \\
NCRES  & 2120     & 199000  &   371  & 8220 \\
NCDIS  & 4146     & 211000  &   776  & 13100 \\ \hline
Total NC  & 7520  & 520000  &   1330 & 26500 \\ \hline
\end{tabular}
\end{center}
\end{table}

In order to allow for a large hydrogen/deuterium target in an underground area, a dedicated facility will be required in order to meet the safety restrictions described above.   This facility will be off-limits to personnel while the detector is filled with either hydrogen or deuterium.  A liquid target is a convenient choice for maximizing the density of the hydrogen or deuterium, which is needed in order to keep the size of the facility small as possible. A target mass of 1~ton will produce a neutrino event sample $100\times$ that of the old bubble chamber data each year in the LBNF beamline.  Table~\ref{tab:fhcH} and Table~\ref{tab:fhcD} show the predicted annual event rates in FHC for hydrogen and deuterium, respectively and Table~\ref{tab:rhcH} and Table~\ref{tab:rhcD} show the predicted event rates in RHC for hydrogen and deuterium.   The flux model is that for the LBNF near site~\cite{DUNE:2020ypp}, and the cross sections are those predicted by GENIE v3~\cite{Andreopoulos:2009rq,Andreopoulos:2015wxa,Tena-Vidal:2021rpu}.  The event yields correspond to an integrated intensity of $1.1 \times 10^{21}$ protons on target (POT), which is expected to be achieved by a 1.2~MW proton beam and a total beamline+detector uptime fraction of 56\%.

\begin{table}[tb]
\begin{center}
\caption{Predicted events per metric ton per year for the LBNF beam ($1.1\times 10^{21}$ POT) on Hydrogen, RHC.  The processes are:  charged-current quasielastic (CCQE), charged-current resonant (CCRES), charged-current deep-inelastic scattering (CCDIS), and their neutral-current counterparts, NCQE, NCRES and NCDIS.}
\vspace{0.3 cm}
\label{tab:rhcH}
\begin{tabular}{|l|r|r|r|r|}
\hline
process &  $\nu_e$ & $\nu_{\mu}$ & ${\bar\nu}_e$ & ${\bar\nu}_{\mu}$ \\ \hline
CCQE      & 0    & 0      & 3200   & 336000 \\
CCRES     & 1750 & 46500  & 2100   & 165000 \\
CCDIS     & 5100 & 107000  & 4800  & 209000 \\ \hline
Total CC  & 6850 & 153000 & 10100  & 710000 \\ \hline\hline
NCQE  & 299      & 9180  & 628  & 65700 \\
NCRES  & 690     & 17800 & 930  & 83100 \\
NCDIS  & 2370    & 50100 & 1480 & 73200 \\ \hline
Total NC  & 3360 & 77100 & 3040 & 222000 \\ \hline
\end{tabular}
\end{center}
\end{table}

\begin{table}[tb]
\begin{center}
\caption{Predicted events per metric ton per year for the LBNF beam ($1.1\times 10^{21}$ POT) on Deuterium, RHC.  The processes are:  charged-current quasielastic (CCQE), charged-current resonant (CCRES), charged-current deep-inelastic scattering (CCDIS), and their neutral-current counterparts, NCQE, NCRES and NCDIS.}
\vspace{0.3 cm}
\label{tab:rhcD}
\begin{tabular}{|l|r|r|r|r|}
\hline
process &  $\nu_e$ & $\nu_{\mu}$ & ${\bar\nu}_e$ & ${\bar\nu}_{\mu}$ \\ \hline
CCQE      & 762    & 23000  & 1160 & 117000 \\
CCRES     & 1870   & 47900  & 2120 & 176000 \\
CCDIS     & 7670   & 158000 & 3570 & 154000 \\ \hline
Total CC  & 10300  & 229000 & 6850 & 447000 \\ \hline\hline
NCQE      & 297  & 9090  & 608  & 63800 \\
NCRES     & 720  & 18600 & 939  & 83200 \\
NCDIS     & 2460 & 52000 & 1410 & 67000 \\ \hline
Total NC  & 3480 & 79700 & 2960 & 214000 \\ \hline
\end{tabular}
\end{center}
\end{table}

Two choices exist for an active-target liquid hydrogen detector -- a TPC that collects drifting electrons from ionized hydrogen molecules and a liquid hydrogen bubble chamber that is read out optically on each spill.  Both a TPC and bubble chamber would need auxiliary devices and detectors to perform high-precision cross section studies.
Both detectors would require a strong magnet for charge separation as well as the measurement of charged particle momentum.
Depending on their size and shape, both detectors could contribute complementary calorimetric measurements for short-lived particles, but heavier particles and those with higher lifetimes would require calorimetry and a muon chamber.

 A liquid hydrogen TPC presents several challenges in design, construction and operation.  The very slow electron drift velocity in liquid hydrogen~\cite{SAKAI198289,HARRISON1971418} presents pileup and background issues, which will need to be evaluated in the context of detector design.   Furthermore, extremely long free-electron lifetimes (of order seconds) will be required in order to collect charge after a long drift distance.  A time-projection chamber can be built with crossed wire planes similar to the single-phase liquid-argon TPC design of the DUNE Far Detector, or it could have pixel readout, similar to the LArTPC near detector design. This would yield a resolution of around 4~mm \cite{DUNE:2021tad}. The presence of high-voltage electrodes in contact with hydrogen will require careful attention to safety, such as the removal of all oxidizers in the hall.

\begin{figure}[t]
    \centering
    \includegraphics[width=1.0\textwidth]{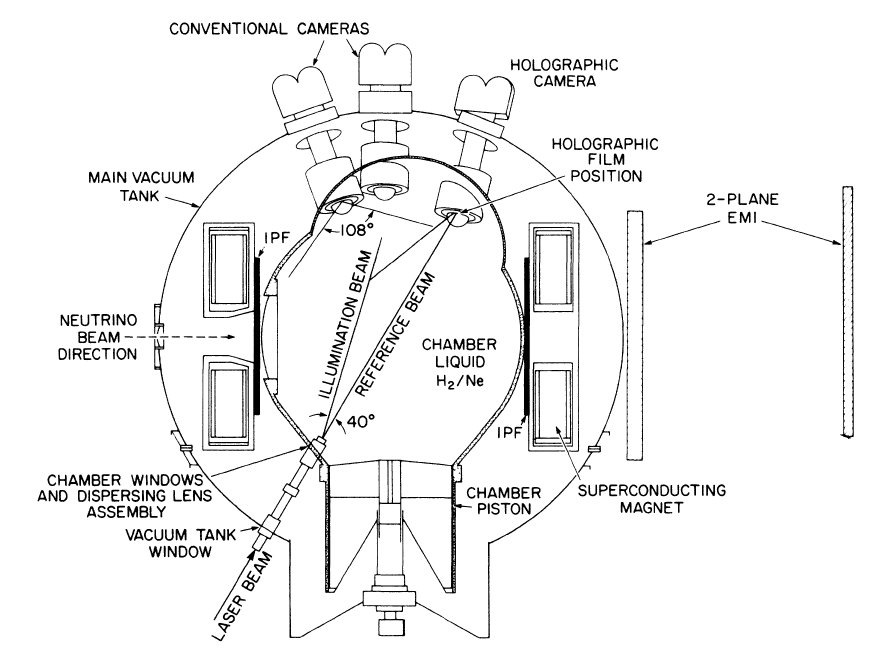}
    \caption{Line drawing of the Fermilab 15 foot bubble chamber.}
    \label{fig:15foot}
\end{figure}

While bubble chambers have many similar safety concerns, they offer several advantages over a similarly-sized hydrogen TPC and have a strong history of successful use in particle physics.
Spatial resolutions of about 100 microns and momentum resolutions of 2\% were standard for historic devices using traditional cameras and relatively weak magnets \cite{Bradner:1960iok}. 
Also, the use of special holographic optical techniques can reduce spatial resolutions down to less than 10 microns \cite{Dykes:1981dvb}. 
An operational gap exists between large, historic bubble chambers like the Fermilab 15-foot bubble chamber and smaller, contemporary, dark-matter-focused chambers like the Scintillating Bubble Chamber \cite{SBC:2021yal}.
This gap will need to be navigated in order to produce a device for LBNF beam.
Historic chambers could cycle quickly ($\sim$1~Hz), but they had relatively short lifetimes which were limited by chamber size, cooling capacity, and the dynamics of bubble nucleation \cite{Baltay:1976okl}.
A concrete example of a historic chamber is the 15-foot bubble chamber that was operated at Fermilab for many years.  Figure~\ref{fig:15foot} shows a line drawing of this chamber, the magnet, the cameras, the piston and the surrounding vacuum tank.   The liquid volume is 30~m$^3$.  We believe that a fiducial volume of $\simeq$ 15~m$^3$ is obtainable and this would have a fiducial mass of roughly 1~ton.

A concern in operating a hydrogen target detector underground is the copious background from muons emerging from the surrounding rock, created by $\nu_\mu$CC interactions.  A preliminary estimate based on the rock-muon flux at the ND hall is that six muons per spill per square meter will emerge from the rock from these interactions assuming a beam power of 1.2~MW~\cite{DUNE:2021tad,ndtf,tmprivate}.  For a 15-foot-diameter circular aperture, this amounts to 120 muons per spill.  If the new hall is located closer to the target, then the rate will be higher due to beam divergence.
Simulations will have to be carried out to estimate the required spatial and/or timing resolution in order to disentangle the $\nu$H interactions in the liquid from muons and other particles, including neutral particles, streaming through the detector coincident with the beam.  An elongated, cylindrical geometry, oriented axially along the beam direction, can help improve the signal-to-background ratio.

\begin{figure}[ht]
    \centering
    \includegraphics[width=1.0\textwidth]{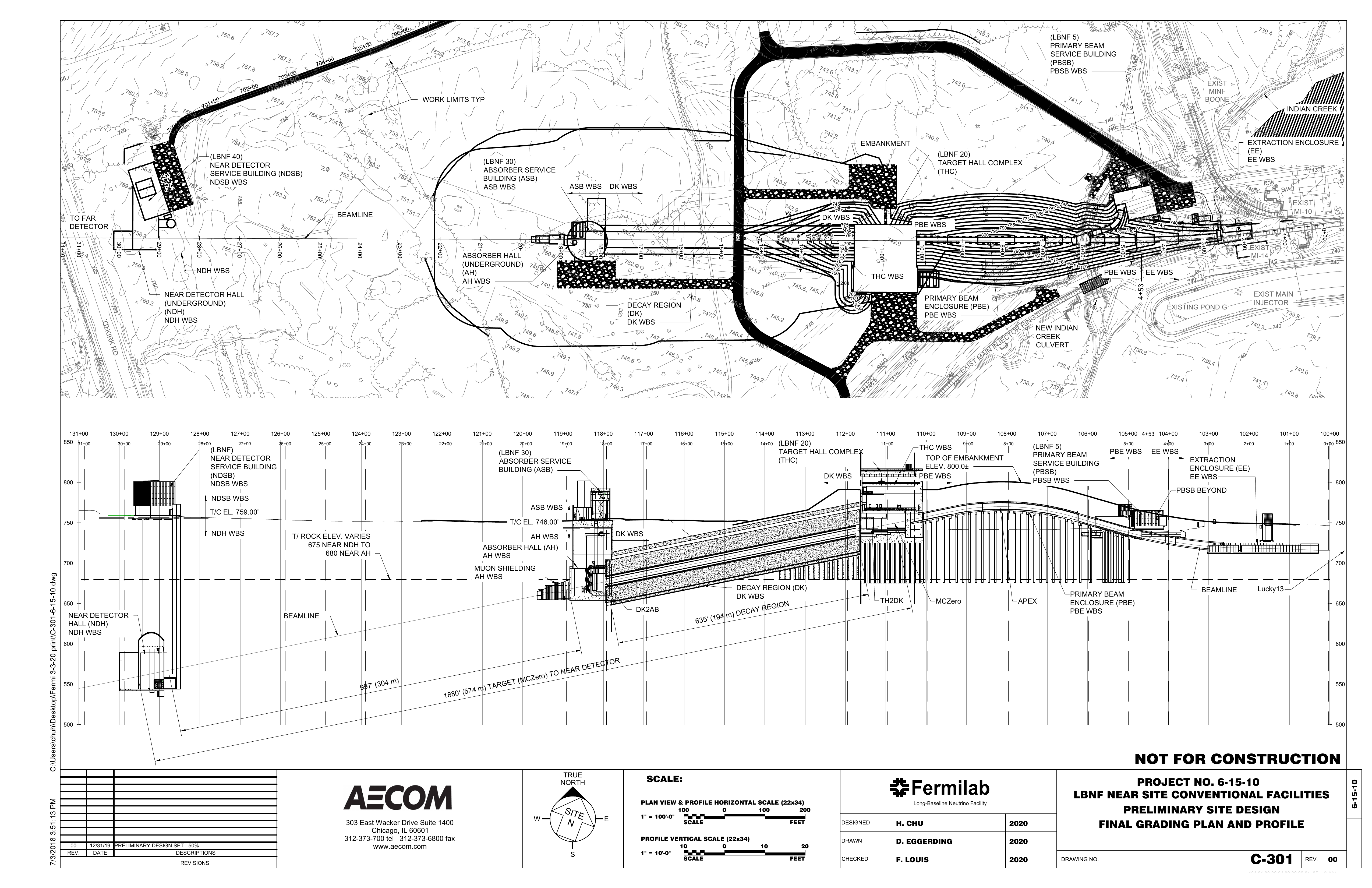}
    \caption{
    Elevation of the LBNF near site, showing the target hall, decay pipe, and the near detector hall.  The boundary between soil and rock is shown.  The vertical scale is expanded by a factor of two with respect to the horizontal scale.  Both scales are measured in feet.}
    \label{fig:lbnfsiteelevation}
\end{figure}

Like detector shape, the location of the new hall is also an optimizable parameter. Figure~\ref{fig:lbnfsiteelevation} shows plan and elevation views of the LBNF near site.  If the hydrogen/deuterium detector is located close to the near detector hall, the flux is more similar to that measured by the near detectors, requiring less extrapolation based on beam models, when comparing the data or using them in combined fits.  A location further upstream may suffer from muons that have not yet ranged out.  Currently, we believe that the beam muon rate at a position 374~m from the target (vs. the 574~m of the near detector hall) is acceptable.  A location further upstream involves less excavation, and also the construction will be less disruptive to the near detectors which may be operating during construction of the new shaft. The cryostat holding the liquid hydrogen or liquid deuterium would be lowered into the shaft at the dedicated facility.  Assembly and testing of the detector will take place in a surface building with access to the shaft connecting to the underground hall.  A mechanism sufficient to raise and lower the detector in the shaft would be required.  

\subsection{Detectors for the Booster Neutrino Beamline and the NuMI Beamline}

Utilizing the existing Booster Neutrino Beamline (BNB) or NuMI beamline for a hydrogen device has multiple advantages
over a dedicated facility in the LBNF beamline.  Since the beamlines already exist and are operating, a hydrogen experiment is not limited in the startup time by the beam construction, although it needs to finish making measurements when the beamlines are decommissioned. There is also a possibility for the continuation of the BNB during the operation of DUNE as the Fermilab Booster will be necessary to supply the LBNF. Another advantage is the relative shallowness of the BNB beamline compared to the LBNF beamline, and the presence of the NuMI hall.  The same safety concerns about flammable gases and liquids mentioned above for an underground experiment in the LBNF beamline apply also to underground apparatus in the NuMI beamline.  The BNB is close enough to the surface that the experimental hall can be on the surface with better ventilation than an underground hall.

The neutrino beam spectrum from the BNB peaks at a lower energy than the LBNF spectrum, and the NuMI spectrum
peaks higher.  Placing a detector off axis reduces the average energy and narrows the spectrum.
Additionally, the timing structure of the NuMI beam consists of short spills of $\approx 10\mu$s in length, separated by 1.2~s.
This timing is ideal for a bubble chamber, which requires significant recovery time after expansion, in order to get ready for the next spill.  The timing structure of the BNB however is more awkward for traditional bubble chambers, with a spill frequency of 15~Hz. Currently, a Lab Directed Research \& Development project has been funded at Fermilab for the development of a modular hydrogen bubble chamber which could be used in LBNF Beam, NuMI Beam, or the BNB.

\subsection{Spin-polarized targets}

Neutrino-nucleon and neutrino-nucleus cross sections are expected to exhibit strong dependences~\cite{PhysRevD.101.073002} on the spins of the initial-state nucleons. The spin asymmetries depend strongly on the nucleon axial form factor with enhanced sensitivity at GeV energies in the antineutrino mode~\cite{PhysRevD.101.073002,Tomalak:2020zlv}.  They are also sensitive to nonstandard interactions.  Targets with spin-polarized nuclei have been in use in laboratories since the 1960s, in charged-particle beams and in beams of photons, but the targets have all been smaller than is needed for a neutrino experiment.  We investigate here the possibility of measuring neutrino interactions on polarized nuclei and discuss expected requirements. We then discuss possible technologies that can meet these requirements.

\subsubsection{Target and detector requirements}
\label{sec:polarized_target_requirements}

\begin{enumerate}
    \item  The target material must be highly polarized.  The degree of polarization of the target $p$ is a dilution factor on measured asymmetries.  A low polarization fraction of the target requires an amount of data proportional to $1/p^2$ to obtain a statistical precision on an asymmetry, relative to a measurement with a 100\% polarized target. The effects of time-dependent drifts in the polarization are also suppressed by the magnitude of the polarization. Targets with heavy nuclei have small polarization fractions due to the fact that most nucleons are in pairs.
    \item  The polarization of the target must be measured with an uncertainty that does not dominate the error budget.
    \item The target must be massive enough to produce an event rate high enough to make interesting measurements.  While ton-sale polarized targets may be very ambitious, a target mass of hundreds of kilograms may be necessary.  The product of the target mass, the beam intensity, and the beam+detector uptime is the relevant figure of merit.  Studies are planned to investigate how many interactions need to be collected.
     \item The detector must be integrated with the target.  This requirement differs from that of charged-particle beam experiments because of the much smaller interaction cross section of neutrinos on nuclei, compared to charged-particle cross sections, and because neutrino beams are more diffuse than focused charged-particle beams. A detector with sufficient mass to produce interesting numbers of neutrino scatters is also large enough to range out the particles produced in these scatters, so particles traversing the target need to be detected.  This can be accomplished by either interleaving the target and detector materials as in MINOS, or by using the target itself as the ionization medium as in a liquid argon TPC.
    \item The active detector must withstand a high magnetic field (2 to 5 Tesla), low temperatures (mK), and applied RF fields that are required to polarize the target and maintain its polarization.
    \item The polarization direction needs to be adjustable.  Measurements of asymmetries require the measurement of cross sections when the target spin is pointing along the beam or in the opposite direction.  Polarization directions transverse to the beam axis are also desirable, but the priority is on longitudinal polarization.   The construction of the target/detector is likely to include a strong solenoidal magnet in a cryostat, giving two natural choices for the polarization direction.
    \item The detector must have a spatial resolution necessary to count particles and measure their angles and momenta and particle ID, when their tracks have very small radii of curvature due to the strong applied magnetic field necessary for maintaining the target polarization.  A downstream muon spectrometer will be needed to contain the higher-energy muons escaping the target.  A calorimeter will be needed to measure photons from $\pi^0$ decay and electrons.
    \item The time scales for cooling the target and polarizing it must be commensurate with the operation plan of the beam.  Maintenance during the annual LBNF beam shutdown is expected, and the amount of time to get the apparatus ready for data taking each year must be a small number of months or less.
\end{enumerate}

The very intense LBNF neutrino beam provides opportunities for even small targets to produce enough data over time to make an impact in this previously unexplored domain.

\subsubsection{Possible technologies}

The requirement of having both high
nuclear target polarization and high heat load simultaneously is achievable through dynamic nuclear polarization (DNP) of solid-state targets.  These systems contain a high-powered microwave generator, a superconducting magnet ($\sim$5T) to produce the Zeeman splitting, and a high-cooling-power evaporation refrigerator used to hold the target at 1~K despite the heat load from the beam, the microwaves, the integrated particle detection, and external heat leaks~\cite{Crabb:1997cy,Dael:1992xh}.

The solid-state polarized target invention in the 1960s leveraged DNP to achieve high polarization of nuclear fixed targets and initiated an era of spin sensitive particle interaction experiments. The very first experiment using a polarized target system was at Saclay~\cite{ABRAGAM1962310}, which was subsequently followed by Lawrence Radiation Laboratory Berkeley~\cite{CHAMBERLAIN1963293}.  At this stage almost every particle physics laboratory around the world have used such target systems at one time or another. The technology of these targets is quite mature and it is now possible to polarize nucleons and other nuclei in any orientation required for beams of photons, muons, electrons, protons, and mesons.  

The heat load to the target and refrigerator are frequently the determining factors of what target system will be required.  A $^3$He or $^4$He evaporation refrigerator is required when there is high beam intensity resulting in a larger heat load (around $\sim$0.1 W or more) to the target cryostat.  At a high enough intensity even a photon beam \cite{Day:2019qdz} requires an evaporation refrigerator.  An evaporation refrigerator (0.5-1~K) receives DNP microwave irradiation continuously to enhance the target polarization over the course of the experiment while sitting in a high polarizing magnetic field (2.5-5 T) with a homogeneity of around $10^{-4}$.

For lower heat loads a $^3$He/$^4$He dilution refrigerator is used where the DNP process nominally takes place between 300-500 mK. A dilution refrigerator DNP system usually has the advantage of running at lower temperature (below 100 mK) once the target has been maximally polarized.  This reduces the spin-relaxation rates so that they are frozen such that the depolarization of the target takes very long compared to the time of the experiment.  Such a system is referred to as a Frozen Spin Target \cite{NIINIKOSKI197195,Isagawa:1977vq,Niinikoski:1981kc,Bradtke:1999zg,Keith:2012ad,frost3,Ball:2003vb} and has the advantage of not requiring continuous DNP so that both the microwaves and the large polarizing magnet (2.5-5 T) can be removed.  Often a holding coil of 0.5~T is positioned to help preserve the frozen spin state during running.  

Due to the negligible heat load from the high-intensity neutrino beam, both the evaporation and frozen spin target could be used.  Both of these systems can also be scaled up longitudinally along the beam line to increase the exposure.  There are, however, limits to the width of the target as both require a strong polarizing magnetic field with high field homogeneity.  It is also possible to increase the target's overall diameter as well but the limiting factor is the size of bore of the magnet~\cite{Berryhill:2019gan} and the expense of building a polarizing magnet on the desired scale (more than a meter).  Polarizing solenoids can be built to have much larger warm bores but to keep the desired homogeneity usually the field strength will need to decrease to keep the magnet construction expenses reasonable. Using a 1.5~T magnet to polarize materials such as NH$_3$ is still practical assuming that the nuclei are polarized at a much lower temperature. The DNP process will then take significantly longer to reach optimal polarization enhancement.

The detector must be integrated with the target due to the low energies of the outgoing production particles and the very small interaction cross section of neutrinos on nuclei.  Historically this type of constraint has been addressed using active targets where the target material also acts as a detector of the low-energy decay products.  For nitroxyl radicals such as TEMPO, DTBN and oxo-TEMPO, the unpaired electron is localized predominately in the N-O bond and is surrounded and shielded by four methyl groups \cite{BUNYATOVA200422}.  This molecular structure allows the dopant to be chemically mixed without losing the free electron.  These free radicals can be combined with materials like polymethyl methacrylate (PMMA) and polystyrene which can also be used as scintillating detectors.  Such approaches provide active targets for experiments that require a polarized target combined with particle detection very close to the polarized target nucleus \cite{BUNYATOVA200422,vandenBrandt:2002ab,vandenBrandt:2002rs}.

Our initial design consists of a scintillator mesh with 1~mm strips running both longitudinally and transversely to provide pixel and cluster reconstruction information.  These strips are connected to wavelength shifting fiber inside the mixing chamber volume of a large scale $^3$He/$^4$He dilution refrigerator. In the insulating vacuum space, these fibers interface and optically connect to a set of Multi-Pixel Photon Counter (MPPC), also known as silicon photomultiplier (SiPM), a solid state photomultiplier comprised of a high density matrix of Geiger-mode-operated avalanche photodiodes.  Routing the optical fibers outside of the mK volume distances the heat source of the SiPMs and associated electronics from the cold target.

The suggested system could be built to house a target with diameter of 25~cm with a length of several meters.  The system could be polarized with a long solenoid magnet (2.5~T) that could move back and forth to polarize different sections and the two different spin states of the long sections of target.  A separate large Helmholtz holding coil (0.3~T) could be positioned outside of the cryostat on the section not being actively polarized with the solenoid. This holding field would preserve the polarization during the frozen spin state.  In this system, the dilution refrigerator is static but the magnets can move on a track in the longitudinal direction to polarize different sections of the target.

\subsubsection{Possible target and detector combinations for a neutrino experiment}
\label{sec:polarized_target_options}

Two fundamentally different approaches can address the main challenges of building a polarized-target neutrino experiment.  A single, large cryostat in a strong magnetic field can provide the necessary environment for the target and the integrated detector, or a modular approach with several smaller cryostats and magnets may be considered.  One example of a large cryostat cooled with a dilution refrigerator is the CUORE dark-matter experiment~\cite{Ouellet:2014qua,Schaeffer_2009}.  This cryostat provides a cubic-meter-scale volume at 10~mK, in which a ton-scale detector made of TeO$_2$ crystals is installed.  In either case, auxiliary detectors to measure neutral particles and escaping muons will be required.

\section{Conclusion \label{sec:summ}}

New measurements of neutrino cross sections on hydrogen and deuterium will have profound impacts on the ability to extract precision results from future neutrino experiments.  Such new measurements will help elucidate the structure of nucleons, provide inputs to precision tests of the Standard Model beyond neutrino oscillations, and provide unique windows on direct searches for new particles and forces.

Neutrino scattering probes nucleon structure
in a way that is not accessible via electron scattering.  
More powerful neutrino beams and larger datasets have the potential to usher in
a new precision era for a range of interesting Standard Model cross sections, 
including quasielastic scattering; production of additional particles such as photons, pions, kaons and hyperons; and inclusive inelastic scattering. 
All of these can be studied with much better accuracy than with historical bubble-chamber experiments.
Scattering of neutrinos on polarized nucleons is characterized by large asymmetries and can provide a powerful new window on nucleon axial structure. 

A variety of BSM signatures can be probed with a hydrogen/deuterium target.  We discussed cases where H/D targets can provide superior sensitivity compared to nuclear target experiments, owing to small hadronic/nuclear uncertainty on interaction cross sections and reconstructed energies. 
Scenarios include cases of BSM particles produced in the production target, produced in meson or muon decays, or produced in neutrino interactions with the hydrogen or deuterium.   
BSM physics can also manifest itself as modified neutrino cross sections.

Elementary form factors can also be calculated using lattice QCD.  If the confidence in the lattice calculations is high, 
they can be combined with experimental constraints on the form factors as inputs to models of neutrino scattering on the heavier nuclei used in neutrino oscillation measurements.  
Alternatively, the lattice QCD predictions can be confronted with future precision experimental data and differences can be used to test for models of new physics and to experimentally measure nonperturbative contributions to radiative corrections. 
Additional constraints on elementary form factors come from muon capture, electron and positron scattering from H and D, and pion electroproduction.  Any deviations from a consistent description of nucleon form factors provides an opportunity to learn about new physics or about the techniques used to predict or measure the form factors.

Experimental options include using the SAND detector's hydrocarbon (CH$_2$)$^n$ radiators as a source of hydrogen.  Pure carbon radiators and precise primary vertex location will allow the separate measurement of neutrino interactions on the hydrocarbon and carbon.  Subtraction of the carbon measurements from the hydrocarbon measurements in the same beam yields estimates of the interaction rates on pure hydrogen.  The use of transverse kinematic imbalance (TKI) variables further improves the signal-to-background ratio before the subtraction and allows for sideband control of important backgrounds.  Unfortunately, the measurements cannot be extended to deuterium in this way as deuterated plastics are prohibitively expensive.

Another option is to include a hydrogen-rich gas in the ND-GAr subdetector.  Hydrogen itself cannot be used for safety reasons, nor even a flammable mixture of hydrogen and other gases.  TKI can help in purifying a sample of reactions on hydrogen when the gas contains molecules with heavier atomic species present.

A separate hall can be constructed upstream of the DUNE Near Detector hall to house a large, traditional bubble chamber and ancillary detectors, such as calorimeters and muon chambers.  It would have to be located at least 150 feet below grade in order to be located on the axis of the LBNF neutrino beam. This option is quite expensive, as it involves constructing everything to an explosion-proof standard, and personnel would have to be excluded from the hall while hydrogen is present.  Placing such a detector in the NuMI underground hall is prohibitive from a safety standpoint.  Placing it on the surface in the BNB beam is more promising; however,
the energy and timing structure of the BNB beam is less optimal for the operation of a traditional bubble chamber. Contemporary bubble chambers designed to study dark matter have long sensitive times and could provide an avenue for taking bubble-chamber data in the BNB beam, which has a 15 Hz spill rate. A Fermilab LDRD project has been commissioned to develop a hydrogen chamber with variable cycling and sensitivity times.

The prospect of measuring neutrino cross sections on a polarized target is very exciting.  It has never been done before, and the inherent polarization of neutrinos in the Standard Model and the parity-violating nature of the weak interactions result in very large polarization asymmetries.  A polarized target would have to be scaled up in size relative to existing ones in order to provide event rates high enough to make useful measurements.  A polarized target would require very low temperatures, of order 1~K, a strong magnetic field, and RF applied so that dynamical nuclear polarization can be used.

Neutrino scattering on hydrogen and deuterium 
probes the fundamental interactions of neutrinos with individual nucleons.  The sparseness of existing data is a limiting factor in neutrino oscillation studies, precision measurements in the Standard Model, and searches for new physics. 
The large investment in the intense neutrino beamlines at Fermilab provides a unique opportunity to 
measure these interactions with an unprecedented level of precision. 

\section*{Acknowledgments}

Work supported by the Fermi National Accelerator Laboratory, managed and operated by Fermi Research Alliance, LLC under Contract No. DE-AC02-07CH11359 with the U.S. Department of Energy. The work is partially supported by the US Department of Energy through the Los Alamos National Laboratory. Los Alamos National Laboratory is operated by Triad National Security, LLC, for the National Nuclear Security Administration of U.S. Department of Energy (Contract No. 89233218CNA000001).
Research of R.J.H., R.P. and O.T. supported by the U.S. Department of Energy, Office of Science, Office of High Energy Physics, under Award No. DE-SC0019095. 
A.S.M. is supported by the Department of Energy, Office of Nuclear Physics, under Contract No. DE-SC00046548.
The work of H.L. is partially supported by the US National Science Foundation under grant PHY~1653405. The Work of O.T. is funded in part by LANL’s Laboratory Directed Research and Development (LDRD/PRD) program under project number 20210968PRD4 and in part by the National Science Foundation under Grant No. NSF PHY-1748958.  The work of X.L. is supported by the Science and Technology Facilities Council (UK). 
The work of C. Wilkinson and B. Russell is supported by the Director, Office of Science, Office of Basic Energy Sciences, of the U.S. Department of Energy under Contract No. DE-AC02-05CH11231. D.G.D. acknowledges Ram\'on y Cajal program (Spain) under contract number RYC-2015-18820. L.A.R. is  supported by the Spanish Ministerio de Ciencia e Innovaci\'on under contract PID2020-112777GB-I00, the EU STRONG-2020 project under the program H2020-INFRAIA-2018-1, grant agreement no. 824093 and by Generalitat Valenciana under contract PROMETEO/2020/023.  M.G.’s work was supported by the EU Horizon 2020 research and innovation programme, STRONG-
2020 project under grant agreement No 824093, and by the Deutsche Forschungsgemeinschaft under the personal grant GO 2604/3-1.  R.P. was supported by the Fermilab Intensity Frontier Fellowship during part of this work. R.P. would like to thank Matheus Hostert for helpful discussions. 

\printbibliography

\end{document}